\documentclass[11pt]{article}
\usepackage{natbib}
\usepackage[dvips]{graphicx}
\begin{document}

\def\lesssim{\mathrel{\hbox{\rlap{\hbox{\lower4pt\hbox{$\sim$}}}\hbox{$<$}}}}
\def\gtrsim{\mathrel{\hbox{\rlap{\hbox{\lower4pt\hbox{$\sim$}}}\hbox{$>$}}}}
\newcommand{\mincir}{\raise
-2.truept\hbox{\rlap{\hbox{$\sim$}}\raise5.truept
\hbox{$<$}\ }}
\newcommand{\magcir}{\raise
-2.truept\hbox{\rlap{\hbox{$\sim$}}\raise5.truept
\hbox{$>$}\ }}
\newcommand{\be}{\begin{equation}}
\newcommand{\ee}{\end{equation}}
\newcommand{\ba}{\begin{eqnarray}}
\newcommand{\ea}{\end{eqnarray}}
\newcommand {\h} {$\mathrm{h^{-1}}$ Mpc $ \;$}
\newcommand {\kpc} {$\mathrm{h^{-1}}$ kpc}
\newcommand {\hh} {$\mathrm{h^{-1}}$ Mpc}
\newcommand {\ks} {km~s$^{-1} \;$}
\newcommand {\kss} {km~s$^{-1}$}
\newcommand {\msun} {$\mathrm{h^{-1}_{70} \  M_{\odot} \;}$}
\newcommand {\m} {$\mathrm{M_{\odot} \;}$}
\newcommand {\ml} {$\mathrm{h \, M_{\odot}/L_{\odot} \;}$}
\newcommand {\mll} {$\mathrm{h \, MSq_{\odot}/L_{\odot}}$}
\newcommand{\vel}{\,{\rm km\,s^{-1}}}
\newcommand{\kms}{\,\mbox{km s$^{-1}$}}
\newcommand{\OI}{[{O}{I}]}
\newcommand{\Lsun}{L$_{\odot}$}
\newcommand{\Msun}{M$_{\odot}$}
\newcommand{\Zsun}{Z$_{\odot}$}
\newcommand{\OIIIOII}{[{O}{III}]/[{O}{II}]}
\newcommand{\OIII}{[{O}{III}]}
\newcommand{\SII}{[{S}{II}]}
\newcommand{\NII}{[{N}{II}]}
\newcommand{\NI}{[{N}{I}]}
\newcommand{\OII}{[{O}{II}]}
\newcommand{\OH}{log(O/H)+12}
\newcommand{\Ha}{H$\alpha\,$}
\newcommand{\Hb}{H$\beta\,$}
\newcommand{\Hd}{H$\delta\,$}
\newcommand{\HeII}{[\ion{He}{2}]}
\newcommand{\NeIII}{[\ion{Ne}{3}]}
\newcommand{\CIII}{[\ion{C}{3}]}
\newcommand{\CIV}{[\ion{C}{4}]}
\newcommand{\SiIII}{[\ion{Si}{3}]}
\newcommand{\HII}{{H}{II}}
\newcommand{\NIII}{[\ion{N}{3}]}
\newcommand{\HeIIHb}{[\ion{He}{2}]$\,\lambda 4686/{\rm H}\beta\,$}
\newcommand{\LIR}{${\rm L_{IR}}\,$}
\newcommand{\LFIR}{${\rm L_{FIR}}\,$}
\newcommand{\bfLIR}{${\rm\bf L_{IR}}\,$}
\newcommand{\NIIHa}{[{N}{II}]/H$\alpha$}
\newcommand{\SIIHa}{[{S}{II}]/H$\alpha$}
\newcommand{\OIHa}{[{O}{I}]/H$\alpha$}

\begin{center}
{\Large ACCESS - V. Dissecting ram-pressure
  stripping \\ through integral-field spectroscopy and multi-band
  imaging}
\bigskip

P. Merluzzi$^{1}$, G. Busarello$^{1}$, M. A. Dopita$^{2,3}$,
  C. P. Haines$^{4,5}$, D. Steinhauser$^6$, A. Mercurio$^{1}$,
  A. Rifatto$^{1}$, R. J. Smith$^{7}$, S. Schindler$^6$\\
  merluzzi@na.astro.it \\ $^1$ INAF-Osservatorio Astronomico di
  Capodimonte, Via Moiariello 16 I-80131 Napoli, Italy\\ $^2$ Research
  School of Astronomy and Astrophysics, Australian National
  University, Cotter Rd., Weston ACT 2611, Australia \\ $^3$ Astronomy
  Department, Faculty of Science, King Abdulaziz University, PO Box
  80203, Jeddah, Saudi Arabia \\ $^4$ School of Physics and Astronomy,
  University of Birmingham, Birmingham B15 2TT UK \\ $^5$ Steward
  Observatory, University of Arizona, 933 N Cherry Avenue, Tucson, AZ
  85721, USA\\ $^6$ Institute of Astro- and Particle Physics,
  University of Innsbruck, Technikerstr. 25, 6020 Innsbruck, Austria
  \\ $^7$ Department of Physics, University of Durham, Durham DH1 3LE
  UK
\end{center}




\label{firstpage}

\begin{abstract}

We study the case of a bright (L$>$L$^\star$) barred spiral galaxy
  from the rich cluster A\,3558 in the Shapley supercluster core
  ($z$=0.05) undergoing ram-pressure stripping. Integral-field
  spectroscopy with WiFeS at the 2.3m ANU, complemented by imaging in
  ultra violet (GALEX), $B$ and $R$ (ESO 2.2m WFI), H$\alpha$
  (Magellan), $K$ (UKIRT), 24$\mu$m and 70$\mu$m (Spitzer), allows us
  to reveal the impact of ram pressure on the interstellar
  medium. With these data we study in detail the kinematics and the
  physical conditions of the ionized gas and the properties of the
  stellar populations. We observe one-sided extraplanar ionized gas
  along the full extent of the galaxy disc, extending $\sim$13\,kpc in
  projection from it. Narrow-band H$\alpha$ imaging resolves this
  outflow into a complex of knots and filaments, similar to those seen
  in other cluster galaxies undergoing ram-pressure stripping. The gas
  velocity field is complex with the extraplanar gas showing signature
  of rotation, while the stellar velocity field is regular and the
  $K$-band image shows a symmetric stellar distribution. We use
  line-ratio diagnostics to ascertain the origin of the observed
  emission. In all parts of the galaxy, we find a significant
  contribution from shock excitation, as well as emission powered by
  star formation. Shock-ionized gas is associated with the turbulent
  gas outflow and highly attenuated by dust (A$_v$=1.5-2.3\,mag). All
  these findings cover the whole phenomenology of early-stage
  ram-pressure stripping. Intense, highly obscured star formation is
  taking place in the nucleus, probably related to the bar, and in a
  region 12\,kpc South-West from the centre. These two regions account
  for half of the total star formation in the galaxy, which overall
  amounts to 7.2$\pm$2.2\,M$_{\odot}$yr$^{-1}$. In the SW region we
  identify a starburst characterized by a $\sim 5\times$ increase in
  the star-formation rate over the last $\sim$100\,Myr, possibly
  related to the compression of the interstellar gas by the ram
  pressure. The scenario suggested by the observations is supported
  and refined by {\it ad hoc} N-body/hydrodynamical simulations which
  identify a rather narrow temporal range for the onset of
  ram-pressure stripping around t$\sim$60\,Myr ago, and an angle
  between the galaxy rotation axis and the intra-cluster medium wind
  of $\sim 45^\circ$. The ram pressure is therefore acting at an
  intermediate angle between face-on and edge-on. Taking into account
  that the galaxy is found $\sim$1\,Mpc from the cluster centre in a
  relatively low-density region, this study shows that ram-pressure
  stripping still acts efficiently on massive galaxies well outside
  the cluster cores, as also recently observed in the Virgo cluster.
\end{abstract}

Keywords: galaxies: evolution -- galaxies: ISM -- galaxies:
intergalactic medium -- galaxies: clusters: general -- galaxies:
stellar content -- galaxies: clusters: individual: A3558

\begin{twocolumn}
\section{Introduction}
\label{sec:int}

Both the current properties and the past evolution of galaxies are
strongly dependent on their environment (e.g. Blanton et
al. \citeyear{bla05}) and mass (e.g. Baldry et al. 
\citeyear{BBB06}, Haines et al. \citeyear{haines07}). The
environment is a determinant for both the morphology-density
\citep{dressler80,dressler97} and the star formation-density
\citep{BO84,LBD02,KWH04} relations: late-type, blue, star-forming
galaxies are predominant in the field, while early-type, red, passive
galaxies are preferentially found in galaxy clusters. This suggests
that blue galaxies accreted from the field have been transformed into
the passive S0s and dEs found in local clusters. A key question in
galaxy evolution is therefore, how does the harsh cluster environment
act to transform the spiral galaxies accreted from the field into
passive S0s and dEs?

Several mechanisms dependent upon the environment have been proposed
and investigated in detail and all of them serve to kinematically
disturb spiral galaxies and/or deplete their reservoirs of
gas, and so quench star formation. These physical processes include
gravitational and tidal interactions amongst galaxies
\citep{TT72,moore}, between galaxies and the cluster gravitational
field (Byrd \& Valtonen 1990), galaxy mergers \citep{BH91},
group-cluster collisions \citep{B01}, ram-pressure \citep{GG72}
and viscous stripping \citep{N82}, evaporation \citep{CS77} and
`starvation' \citep{LTC80}.

Since these mechanisms are characterized by different time-scales and
efficiencies which depend, in turn, on the properties of both the
galaxies and their environment, they can differentially affect the
galaxy properties (Boselli \& Gavazzi \citeyear{BG06}; Haines et
al. \citeyear{haines07}). Over the last ten years, the development of
huge spectroscopic surveys such as the Sloan Digital Sky Survey
(SDSS), plus the availability of panoramic far ultra-violet -- far
infra-red (FUV--FIR) data from the {\em GALEX} and {\em Spitzer} space
telescopes have allowed the impact of environment on star formation to
be quantified in unprecedented detail. However, because they only
consider the global properties of each galaxy (e.g. stellar mass,
star-formation rate), they are unable to identify which process(es)
are behind these transformations, except {\em indirectly} by
statistical comparison of trends with predictions from cosmological
simulations \citep[e.g.][]{BNM00,HSE09}. 
What is fundamentally required are {\em direct}
``smoking-gun'' observations of galaxies in the process of being
transformed via interaction with their environment. The physical
processes behind this transformation can then be distinguished by {\em
resolving} their impacts on the gas contents, kinematics and star
formation of the galaxies.
 
Ram pressure resulting from the passage of the galaxy through the hot
and dense intra-cluster medium (ICM) can effectively remove the cold
gas supply \citep{GG72,AMB99} and thus rapidly terminate ongoing star
formation in cluster galaxies. This process could explain the lower
star formation rates (SFRs; e.g. Balogh et al. \citeyear{BNM00}) and the
redder colours \citep{BNB09} seen in cluster galaxies with respect to
the field population. Nevertheless, ram-pressure stripping, as
originally proposed by \citet{GG72} requires, in principle, the
presence of a dense ICM. Thus, its evolutionary effect would be
limited to cluster cores where the gas discs of massive spirals are
rapidly truncated.

In the last decade ram-pressure stripping (RPS) has been extensively
studied using hydrodynamical simulations (e.g. Roediger \& Hensler
\citeyear{RH05}; Roedigger \& Brugger \citeyear{RB06}; Kronberger at
al. \citeyear{KKF08a}; Kapferer et al. \citeyear{KSS09}; Tonnesen \&
Bryan \citeyear{TB09}; Bekki \citeyear{B09}; J\'achym et
al. \citeyear{JKP09}) describing the RPS process as function of galaxy
parameters (stellar mass, inclination, orbit, velocity, mass of the
gas halo, structure of the interstellar medium, etc.) and the density
and structure of the ICM, suggesting a number of observables which can
be used as diagnostics of its occurrence, intensity, and stage. With
a 3--D hydrodynamical simulation, \citet{MBD03} found that RPS may
extend to poorer environments for low-mass galaxies which, thanks to
their lower escape velocities, are easier to strip. RPS also directly
affects the density and temperature structure of hot halo gas of
galaxies and is relevant in the starvation mechanism in clusters and
groups \citep{LTC80,BCS02,MFF08}.

The characteristic signatures of RPS are the presence of gas outflows,
distortion and ultimate truncation of the gaseous disc without
corresponding distortion of the old stellar component (e.g. Kenney et
al \citeyear{KvG04}; \citeyear{VSC08}; \citeyear{AKC11}). Such
features have been observed in numerous cluster spirals (e.g. Vogt et
al. \citeyear{VHG04}; Chung et al. \citeyear{CvG09}). Ram pressure can
also compress and shock the interstellar medium (ISM) enhancing the
star formation in the inner disc \citep{BV90,FN99}. Tails of stripped
gas may also be present, whose nature and evolution depend also on the
properties of the ISM \citep{RB08,TB10}.

Considering both the theoretical predictions and the observable
effects, it follows that the detection of the stripped gas is a key
step in order to better understand the role of RPS, but also of tidal
stripping, in the transformation of galaxies from their actively
star-forming phase to their passive phase. The stripped gas in cluster
galaxies has been detected through narrow-band H$\alpha$ imaging
(e.g. Gavazzi et al. \citeyear{GBM01}; Yagi et al. \citeyear{YKY07},
\citeyear{YYK10}; Yoshida et al. \citeyear{YYO02}; Kenney et
al. \citeyear{KTC08}, Smith et al. \citeyear{SLH10}), through HI
imaging (e.g. Oosterloo \& van Gorkom \citeyear{OvG05}; Chung et
al. \citeyear{CGK07}; Haynes et al. \citeyear{HGK07}; Koopmann et
al. \citeyear{KGH08}), in X--ray images (e.g. Irwin \& Sarazin
\citeyear{IS96}; Wang et al. \citeyear{WOL04}; Sun \& Vikhlinin
\citeyear{SV05}; Sun et al. \citeyear{SFN06}; \citeyear{SDR10};
Machacek et al. \citeyear{MJF06}; Kim et al. \citeyear{KKF08}) and
with multi-band observations (e.g. Crowl et al. \citeyear{CKG05};
Cortese et al. \citeyear{CMR07}; Abramson et al. \citeyear{AKC11}).

The presence of gas outside of the galaxy disc could also arise from
mechanisms other than RPS, such as tidal interactions. Multi-band
observations of both the tail and the galaxy are needed to distinguish
amongst the different gaseous components in the tail (atomic,
molecular, ionized), and high resolution observations are required in
order to resolve the physics of both the gas and the stars in the
galaxy.

Integral field spectroscopy (IFS) allows us to resolve the different
spatial components in the galaxies, and so measure the dynamical
disturbance of the stellar component, discover disturbed gas velocity
fields, and determine the local enhancement and spatial trend of the
star formation. Recent IFS observations of individual galaxies
(e.g. Cort\'es et al. \citeyear{CKH06}; Crowl \& Kenney
\citeyear{CK06}; Farage et al. \citeyear{FMD10}; Jim\'enez-Vicente et
al. \citeyear{JVM10}; Rich et al. \citeyear{RDK10}; S\'anchez et
al. \citeyear{SRK11}) and ever larger samples of galaxies (e.g
Monreal-Ibero et al. \citeyear{MAC10}; Crowl \& Kenney
\citeyear{CK08}; Pracy et al. \citeyear{PKC09}, \citeyear{POC12}) have
demonstrated the efficiency of this tool both to investigate the
nature of the galaxy emissions and to understand the essential
physics.

A key aim of the ACCESS project\footnote{The project ACCESS (\emph{``A
Complete CEnsus of Star formation and nuclear activity in the Shapley
supercluster''}), a European International Research Staff Exchange
Scheme of the 7th Framework Programme involving the Italian Institute
for Astronomy and Astrophysics - Astronomical Observatory of
Capodimonte, the Australian National University, the University of
Birmingham and the University of Durham.} (PI P. Merluzzi, see
Merluzzi et al. \citeyear{MMH10}, hereafter Paper\,I) is to detect the
signatures of galaxies caught in the act of transformation, using the
unique combination of a large-scale IFS survey with the Wide Field
Spectrograph (WiFeS) at the Australian National University 2.3m
telescope together with an unprecedented data-set (from
far-ultraviolet to far-infrared) on the galaxies in the core of the
Shapley supercluster (SSC) at $z$=0.048
\citep{sos1,MMH10,HBM11a,HBM11b,HBM11c,smith07}, including new
narrow-band H$\alpha$ imaging obtained with the Maryland-Magellan
Tunable Filter (MMTF) on the Magellan-Baade 6.5m telescope. 

Our ongoing ACCESS IFS survey with WiFeS (Dopita et
al. \citeyear{dopita07}, \citeyear{dopita10}) commenced in April 2009,
just after the WiFeS commissioning. The galaxies observed with WiFeS
sample different environments, from dense cluster cores to the regions
where cluster-cluster interactions are taking place, to the much less
populated areas. The whole sample consists of 24 galaxies assessed as
belonging to the Shapley supercluster according their spectroscopic
redshift, and classified as either star-forming, AGN or composite
galaxies using AAOmega spectroscopy \citep{smith07}. They mostly have
intermediate IR colours (0.15$<$f$_{24{\mu}{\rm m}}$/f$_K{<}1$.0; a
proxy for specific-SFR) between star-forming and passive galaxies
(Haines et a. \citeyear{HBM11b}, hereafter Paper\,III) and $K<13$\,mag
($K<K^\star$+1.3). All the galaxies are well-resolved in the optical
images, and often display either disturbed morphology, such as
asymmetry and tails, or evidence of star-formation knots. These
galaxies have full photometric coverage which provides complementary
information on their star formation. In this work we present the
results for the galaxy -- SOS\,114732, a bright (L$\gtrsim$L$^\star$)
spiral galaxy in the rich cluster A\,3558. 

\begin{figure*}
\includegraphics[width=80mm]{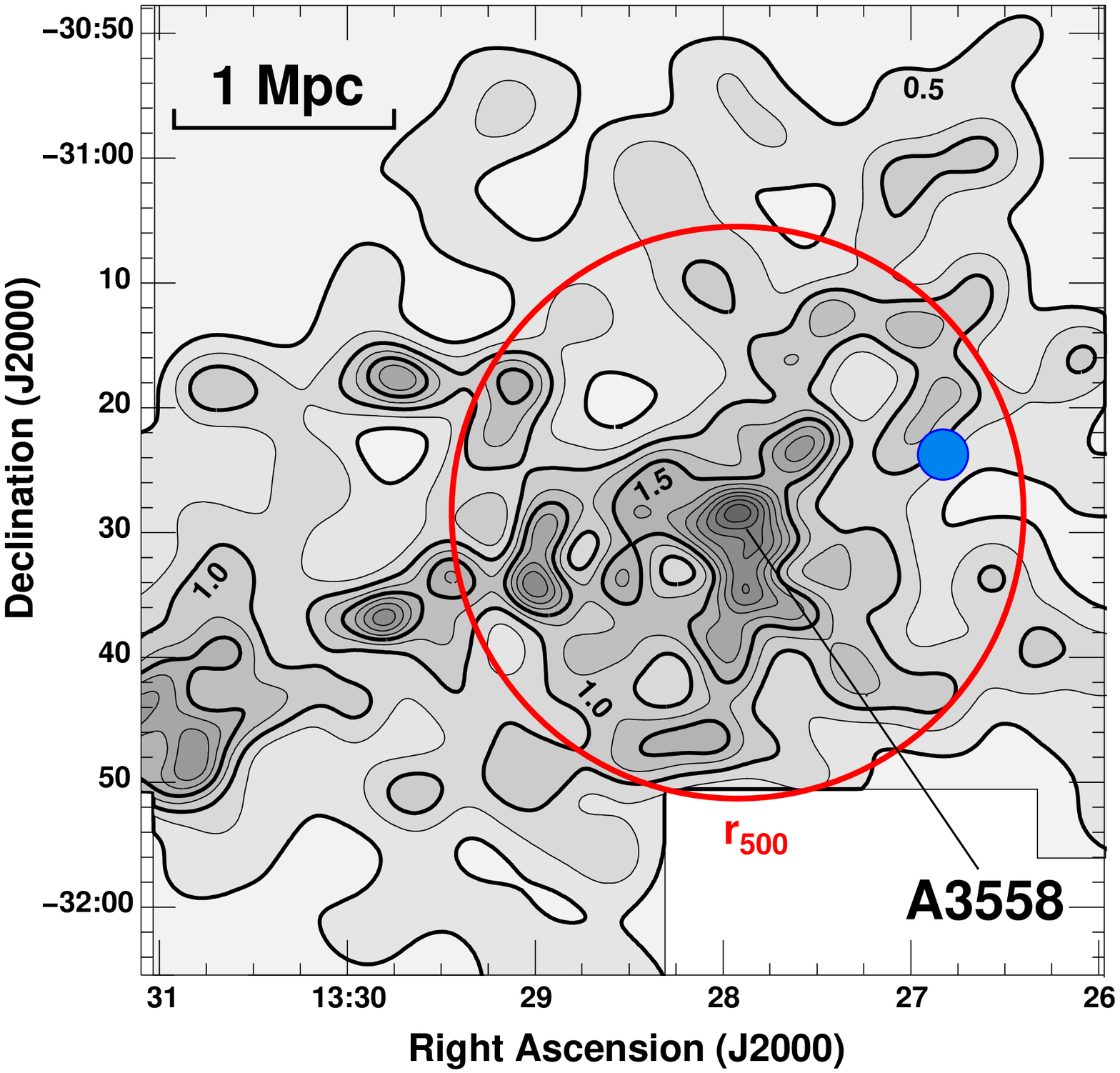}
\includegraphics[width=75mm]{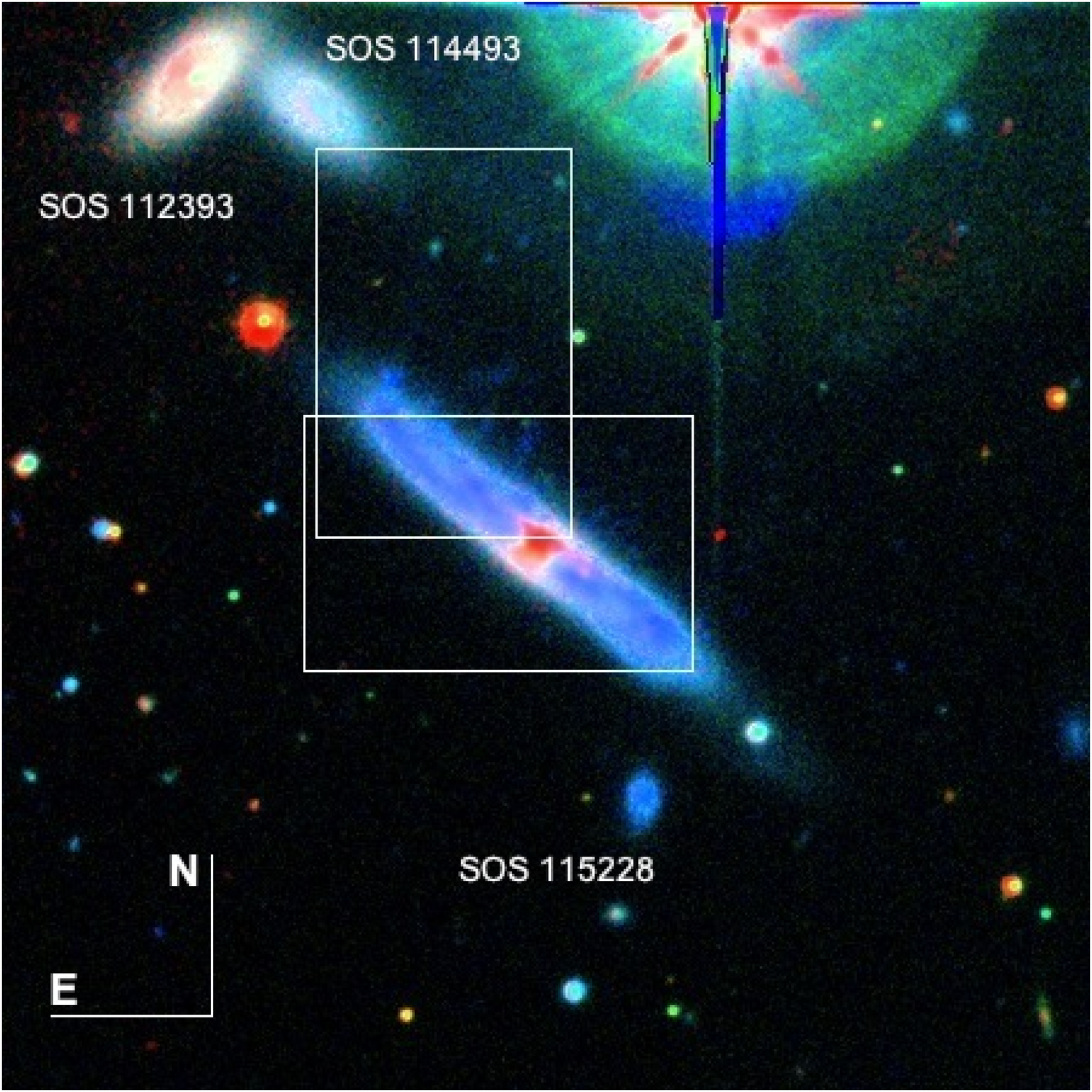}
\caption{{\it Left}: The position of SOS\,114372 is marked (blue dot)
in the field of A\,3558 (North up, East left) where the surface
density of $R{<}2$1.0 galaxies from the SOS is shown in gray scale
(from 0.33 to 2.0 galaxies arcmin$^{-2}$). The $r_{500}$ cluster
radius is indicated by the red circle. {\it Right}: $BRK$ composite
image of SOS\,114372. The two white rectangles denote the WiFeS
pointings: \#1 centred on the galaxy, \#2 covering the NE side. The
two galaxies to the NE are supercluster members: SOS\,114493
($z$=0.05) and SOS\,112393 (z=0.046). SW of SOS\,114372 is a faint
galaxy, SOS\,115228, without redshift measurement. The orientation is
indicated and adopted for all the following galaxy maps. The white
bars indicating the orientation correspond to a length of
15\,kpc. Channels R, G, B are assigned to $K$, $R$, $B$ bands,
respectively. The nuclear region appears to be highly obscured by dust.}
\label{SOS}
\end{figure*}

In Sect.~\ref{gal} we describe the target. The details of IFS
observations and data reduction are given in Sect.~\ref{data} and the
data analysis is described in Sect.~\ref{anl}. The narrow-band
H$\alpha$ imaging obtained using the \\ MMTF are presented in
Sect.~\ref{mmtf}. We present the results in Sect.~\ref{res} where the
morphology of ionized gas, gas and stellar kinematics and the dust
attenuation map are analysed. The physical properties of the gas and
the star formation across the galaxy are discussed in
Sect.~\ref{gas_sf}. All our findings strongly suggest the occurrence
of RPS as outlined in Sect.~\ref{dis} where the nature of the gas
outflow is discussed considering also other possible causes as
galactic winds and tidal interactions. The RPS is confirmed in
Sect.~\ref{simul} by {\it ad hoc} hydrodynamical simulations. We
summarize the results and draw conclusions in Sect.~\ref{con}.

Throughout the paper we adopt a cosmology with $\Omega_M$=0.3,
$\Omega_\Lambda$= 0.7, and H$_0$=70\,km\,s$^{-1}$Mpc$^{-1}$. According
to this cosmology 1\,arcsec corresponds to \\ 0.941\,kpc at $z$=0.048 and
the distance modulus is 36.66.

\begin{table*}
  \centering
\caption{{\bf Properties of the galaxy SOS\,114372.}}
  \begin{tabular}{lcl}
    \hline \hline 
{\bf Property} & {\bf Value} & {\bf Source}\\ 
\hline
Magnitudes and fluxes &&\\
\hline
FUV$^a$ (1510~\AA) & 17.11${\pm}$0.05 & Paper\,II\\ 
NUV$^a$ (2310~\AA)& 16.68${\pm}$0.03 & Paper\,II\\ 
$B^b$ & 15.16${\pm}$0.04 & Mercurio et al. \citeyear{sos1}\\
$R^b$ & 14.27${\pm}$0.04 & Mercurio et al. \citeyear{sos1}\\
$K^b$ & 11.504$\pm$0.015 & Paper\,I\\
24$\mu$m & $42{\pm}2$\,mJy & Paper\,II\\
70$\mu$m & $0.758{\pm}0.038$\,Jy & Paper\,II\\
90$\mu$m & $0.837{\pm}0.086$\,Jy & Murakami et al. \citeyear{MBB07}\\
100$\mu$m & $1.2{\pm}0.2$\,Jy & Allen et al. \citeyear{ANM91}\\
1.4GHz & 9.31\,mJy & Miller \citeyear{M05}\\
\hline
$K$-band bulge-disc decomposition & & \\
\hline
disc radial scale length ($r_d$)& 5.3\,arcsec & this paper\\
bulge effective radius & 1.5\,arcsec & this paper\\
S\'ersic index of the bulge & 1.2 & this paper\\
bulge-to-disc ratio & 0.44 & this paper\\
disc inclination & 82$^\circ$ & this paper\\
\hline
Velocities & & \\
\hline
rotation velocity at $r_d$ & 200$\pm$13\,km\,s$^{-1}$& this paper\\
radial velocity$^c$ & 826$\pm$13\,km\,s$^{-1}$ & Dale et
al. \citeyear{DGH99}; Proust et al. \citeyear{PHC06}\\
\hline
Masses & & \\
\hline
Stellar mass & 7.0${\pm}0.7{\cdot}10^{10}$M$_\odot$ & Paper\,I\\
Dynamical mass$^d$ & $2{\cdot}10^{11}$\,M$_\odot$ & this paper\\
Total halo mass & $1.2{\cdot}10^{12}$\,M$_\odot$ & this paper\\
\hline
Distances & & \\
\hline
Redshift & 0.0506 & Dale et al. \citeyear{DGH99} \\
Projected distance to cluster centre & 0.995\,Mpc & Sanderson \& Ponman \citeyear{SP10}\\
\hline
Star formation rates & & \\
\hline
Global SFR from UV+IR& 8.53$^{+2.55}_{-1.52}$\,M$_\odot$yr$^{-1}$ & Paper\,II\\
Global SFR from H$\alpha$& 7.2$\pm 2.2$\,M$_\odot$yr$^{-1}$ & this paper\\
\hline
\hline
a) AB photometric system. & & \\
b) Vega photometric system. & & \\
c) With respect to the cluster systemic velocity. & & \\
d) Within 20\,kpc radius. & & \\  
\end{tabular}
\label{SOS114372}
\end{table*}

\section[]{The galaxy SOS\,114372}
\label{gal}

The galaxy SOS\,114372, named following the Shapley Optical Survey
(SOS) identification (Mercurio et al. \citeyear{sos1}; Haines et
al. \citeyear{haines06}), is a bright (L$\gtrsim$L$^\star$,
$K$=11.504$\pm$0.015) spiral galaxy at redshift $z$=0.0506
\citep{DGH99} belonging to the rich cluster A\,3558 (Abell richness 4)
which has a median redshift $z$=0.0477 \citep{SHN04,PHC06} and a
velocity dispersion of 1010\,km\,s$^{-1}$ \citep{PHC06}. SOS\,114372
is almost edge-on and is located at the projected distance of
0.995\,Mpc from the centre of the cluster A\,3558, well within the
cluster radius \\ r$_{500}$=1.214$\pm$0.044\,Mpc (see Sanderson \& Ponman
\citeyear{SP10}) and in a relatively low-density region ($\rho$=0.88
gals arcmin$^{-2}$) as measured from the SOS $R<21$ galaxies (Fig.~1,
left panel). This density is less than half of the peak density
observed in the core of A\,3558.

The main properties of the galaxy are listed in Table~\ref{gal}. The
galaxy is amongst the brightest SSC members in the infra-red and it is
also the brightest SSC galaxy in the ultra-violet (UV) (see
Table~\ref{gal} Haines et al. \citeyear{HBM11a}, hereafter Paper\,II),
providing evidence of the presence of both obscured and unobscured
star formation. The total infra-red luminosity of the galaxy is
estimated as $L_{TIR}{=}1.1{\pm}0.1{\cdot}10^{11}L_{\odot}$ based on
fitting the infra-red SED models of \citet{RA09} to the Spitzer/MIPS
70$\mu$m flux and AKARI 90$\mu$m flux \citep{MBB07}. These data are
also consistent with the prior IRAS 100$\mu$m flux measurement
\citep{ANM91}. Miller (\citeyear{M05}) measured a radio flux at 1.4GHz
of 9.31\,mJy corresponding to a luminosity
L$_{1.4GHz}{=}5.3{\cdot}10^{22}$W\,Hz$^{-1}$, consistent with
expectations from the FIR--radio correlation in which both the FIR and
radio emissions come from ongoing star formation
(Paper\,III). Following \citet{leroy}, we estimated a global
SFR=8.53$^{+2.55}_{-1.52}$\,M$_\odot$yr$^{-1}$ (of which 78\% is
obscured, Paper\,II) with the errors accounting for the SFR
calibration uncertainty. The most intense star formation occurs in the
heavily obscured centre, but an intense region of less-obscured star
formation is also observed in the SW disc (see Sect.~\ref{SF}).

We estimate a stellar mass of
$\mathcal{M}_{\star}{=}7.0{\pm}0.7{\cdot}10^{10}$M$_\odot$ (Paper\,I).
SOS\,114372 has dynamical mass within 20\,kpc radius of
$\mathcal{M}_{dyn}{\sim}2{\cdot}10^{11}$\,M$_\odot$ (see
Sect.\ref{GSK}) from which we can estimate the total halo mass
following \citet{CBE05} of
$\mathcal{M}{=}1.2{\cdot}10^{12}$\,M$_\odot$. To derive the
structural properties of the galaxy, we modelled the light
distribution in $K$ band with a S\'ersic profile for the bulge plus an
exponential disc using GALFIT \citep{PHI10}. The choice of the
$K$-band allows to mitigate the effect of strong dust absorption. The
disc radial scale length is $r_d$=6.4 arcsec ($\simeq6$\,kpc) which,
after applying the ``dust correction'' introduced by Graham \& Worley
(\citeyear{GW08}) becomes $r_d\simeq5.3$\,kpc. The effective radius
of the `bulge' is $r_e$=1.5 arcsec ($\simeq1.4$\,kpc). The `bulge' is
significantly elongated, with an axis ratio of 0.7, is tilted with
respect to the disc by 13$^\circ$, and its S\'ersic index is n=1.2
indicating that it is actually a bar. The disc axis ratio is 0.14,
which implies an inclination of the disc with respect to the line of
sight of 82$^\circ$. The $K$-band bulge-to-disc ratio is B/D=0.44,
after correcting for inclination following \citet{DPT08}. 

The $BRK$ composite image, as derived from the SOS (ESO-WFI, Mercurio
et al, \citeyear{sos1}) and $K$-band surveys (UKIRT-WFCAM, Paper\,I),
shows evidence of dust absorption in and SE of the galaxy centre
(Fig.~\ref{SOS}, right panel). In the composite image we also notice
hints of matter beyond the stellar disc (north-west from the disc, see
Sect.~\ref{res}). The orientation of Fig.~\ref{SOS} with North
upwards and East to the left will be adopted throughout the article.

\section[]{Integral-field spectroscopy}
\label{data}

\subsection[]{Observations}
\label{obs}

The spectroscopic data of SOS\,114372 were obtained during two
observing runs in April 2010 (pointing \#1) and April 2011 (pointing
\#2) using the Wide-Field Spectrograph (WiFeS; Dopita et
al. \citeyear{dopita07}, \citeyear{dopita10}) on the Australian
National University 2.3m telescope at the Siding Spring Observatory,
Australia. WiFeS is an image-slicing integral-field spectrograph that
records optical spectra over a contiguous
25$^{\prime\prime}$$\times$\,38$^{\prime\prime}$ field-of-view. This
field is divided into twenty-five 1$^{\prime\prime}$-wide long-slits
(`slices') of 3$8^{\prime\prime}$ length. The spectra were acquired in
`binned mode', with 25$\times$38 spaxels of
1$^{\prime\prime}\times$\,1$^{\prime\prime}$ size. Hereafter we will
refer to the spatial directions across the slices or along them as `X'
or `Y' direction respectively. WiFeS has two independent channels for
the blue and the red wavelength ranges. We used the B3000 and R3000
gratings, allowing simultaneous observations of the spectral range
from $\sim $3300\,\AA\ to $\sim $9300\,\AA\ with an average resolution
of R=2900. For further details on the WiFeS instrument see Dopita et
al. (\citeyear{dopita07}, \citeyear{dopita10}).

The positions of the WiFeS field-of-view for the two pointings are
shown in Fig.~\ref{SOS} (right panel). The total integration time on
the galaxy was of 4.5\,h in run \#1, obtained as the sum of
6$\times$45\,min exposures. Pointing \#2 was obtained to map the gas
distribution in the Northern side of the galaxy and to measure the
extent of extraplanar gas. The data of this pointing are shallower
(2$\times$45\,min) than run \#1 and are used to map the gas and derive
the kinematics, but not to measure line flux ratios. For each galaxy
spectrum, we also acquired the spectrum of a nearby empty sky region
with 22.5\,min exposure to provide accurate sky subtraction.

For each galaxy exposure we obtained spectra of spectrophotometric
standard stars for flux calibration and stars with nearly featureless
spectra to monitor the atmospheric absorption bands. Arc and bias
frames (see below) were also taken for each science exposure.
Internal lamp flat fields and sky flats were taken twice during both
runs.

\subsection{Data reduction}
\label{red}

The data were reduced using the WiFeS data reduction pipeline
\citep{dopita10} and purposely written FORTRAN and
IDL\footnote{http://www.exelisvis.com/language/en-US/ProductsServices/IDL.aspx}
codes. The WiFeS pipeline performs all the steps from bias subtraction
to the production of wavelength- and flux-calibrated data-cubes for
each of the $B$ and $R$ channels. The average r.m.s. scatter around
the dispersion relation was 0.2--0.3\,\AA. The final spectral
resolution achieved is $\sigma{\sim}40$\,km\,s$^{-1}$, and is
wavelength and position dependent (see below). The data-cubes were
sampled at 1$^{\prime\prime}$$\times$1$^{\prime\prime}$$\times$1\,\AA\
and cover a useful wavelength range of 3600--9000\,\AA. The data-cubes
are also corrected for atmospheric differential refraction.

Accurate sky subtraction is particularly important for our data
because of the presence of OH lines in the vicinity of the H$\alpha$
line and the [N{\sc ii}] doublet at the redshift of the SSC. Sky
subtraction was carried out as follows. The sky spectrum taken closest
in time to the galaxy spectrum was first cleaned of cosmic ray hits
using the code LACosmic \citep{vD01} to detect them, followed by an
interpolation along the spatial direction to remove them. The sky
spectra were then smoothed in the Y direction with a median
filter. This allowed us to effectively minimize the introduction of
noise in sky subtraction. In an ideal case, we should have multiplied
the sky cubes by a factor of two (the ratio of exposure times) before
subtraction, but due to differences in airmass and changes in sky
conditions, especially affecting the amplitudes of emission lines, we
normalized the sky frames to the galaxy frames by comparing the
relative strengths of the atmospheric emission lines. This resulted in
a range of multiplicative factors for the sky frames
(${\sim}1$.5--2.5).

Particular attention was devoted to the removal of the atmospheric
absorption features, since at the redshift of the SSC the atmospheric
band at $\sim $6870\,\AA\ may affect the [N{\sc ii}]-H$\alpha$
group. This O$_2$ band is accompanied by two other O$_2$ bands ($\sim$
6280\,\AA\ and $\sim$ 7600\,\AA\ ) which could be effectively used to
monitor the quality of the subtraction.

The co-addition of the individual exposures requires an accurate
spatial registration of the data-cubes. For this purpose, we took
advantage of the high quality ESO-2.2m WFI $B$- and $R$-band images of
the galaxy from the SOS. The bandpasses of these images can be matched
to the $B$ and $R$ channels of WiFeS. The true spatial scale of WiFeS
data was determined from (i) the distance between the galaxy nucleus
and a star taken during run \#1 and (ii) images of the centre of the
globular cluster NGC\,3201 acquired in run \#2 for this purpose. We
find that while the X scale of WiFeS is 1$^{\prime\prime}$/pixel, the
Y scale is larger than the nominal 0.5$^{\prime\prime}$/pixel by a
factor 1.15 for the $B$ data-cube and 1.04 for the $R$
data-cube. After correcting for these pixel scales we multiplied the
WiFeS cubes with the WFI response curves and determined the relative
spatial position by cross-correlation. The individual exposures were
then registered and co-added using IRAF. $B$ and $R$ fluxes were
adjusted, where necessary, by matching the fluxes in the common
wavelength range (5300--5600\,\AA\ ). The adjustment required was
always within the estimated flux errors (see Sect.~\ref{anl1}). The
reduced data-cubes were finally corrected for Galactic extinction
following Schlegel et al. (\citeyear{SFD98}) and using the extinction
curve by \citet{CCM89} with R$_V$=3.1. The sensitivity of our data
turns out to be
0.5$\cdot$10$^{-17}$erg\,s$^{-1}$cm$^{-2}$\AA$^{-1}$\,arcsec$^{-2}$ at
a signal-to-noise ratio SNR=5. This is the lowest SNR for which we
consider our measurements reliable.

\section{H$\alpha$ imaging}
\label{mmtf}

Deep H$\alpha$ imaging of the galaxy SOS\,114372 was obtained with the
Maryland-Magellan Tunable Filter \citep[MMTF;][]{veilleux} on the
Magellan-Baade 6.5m telescope at the Las Campanas Observatory in Chile
on 20 May 2012. The MMTF is based on a Fabry-Perot etalon, which
provides a very narrow transmission bandpass (${\sim}$5--12{\AA}) that
can be tuned to any wavelength over ${\sim}5$000--9200{\AA}
\citep{veilleux}. Coupled with the exquisite image quality provided by
active optics on Magellan and the Inamori-Magellan Areal Camera \&
Spectrograph (IMACS), this instrument is ideal for detecting
extra-galactic H$\alpha$-emitting gas. The MMTF 6815-216
order-blocking filter with central wavelength of 6815{\AA} and FWHM of
216{\AA} was used to provide coverage of the H$\alpha$ emission line
for galaxies belonging to the Shapley supercluster.

At the start of the observing run the etalon plates were parallelized
and the wavelength calibration of the etalon for our setup was
determined and further checked throughout the night. The MMTF does not
provide a monochromatic image over the full extent of the IMACS
$27^{\prime}{\times}27^{\prime}$ field of view, but rather a central
circular monochromatic region (the Jacquinot spot), outside of which
the central transmission wavelength monotonically decreases with
increasing angular distance from the optical axis in a well understood
manner. The instrumental set-up was chosen to provide the largest
Jacquinot spot (${\sim}11.5^{\prime}$) and a transmission bandpass of
FWHM 10.3{\AA} (corresponding to ${\sim}4$70\,km\,s$^{-1}$). The
target galaxy was always placed at the same location, well within the
Jacquinot spot, but away from the gaps between CCDs.
   
The galaxy was observed for a total 60 minutes in H$\alpha$
($4{\times}900$\,sec). The galaxy was also observed for 15 minutes in
the continuum, by shifting the central wavelength of the etalon
${\sim}5$0{\AA} bluewards to exclude emission from both the H$\alpha$
line and the adjacent [N{\sc ii}] lines, and into a wavelength region
devoid of major skylines. The typical image resolution for these
exposures was $0^{\prime\prime}.50$.

These data were fully reduced using the MMTF data reduction
pipeline\footnote{http://www.astro.umd.edu/$\sim$veilleux/mmtf/datared.html},
which performs bias subtraction, flat fielding, sky-line removal,
cosmic-ray removal, astrometric calibration and stacking of multiple
exposures \citep[see][]{veilleux}. Photometric calibration was
performed by comparing the narrow-band fluxes from continuum-dominated
sources with their known $R$-band magnitudes obtained from our
existing WFI images. Conditions were photometric throughout and the
error associated with our absolute photometric calibration is
${\sim}10$\%. The effective bandpass of the Lorenzian profile of the
tunable filter of ${\pi}/2{\times}$FWHM is then used to convert the
observed measurements into $H{\alpha}$ fluxes in units of
erg\,sec$^{-1}$\,cm$^{-2}$. The filter is sufficiently narrow that
there should be little or no contamination from [N{\sc ii}]
emission. The data were obtained in dark time resulting in very low
sky background levels, with 1$\sigma$ surface brightness fluctuations
within a 1\,arcsec diameter aperture of
$0.2{\times}10^{-17}$\,erg\,s$^{-1}$\,cm$^{-2}${\AA}$^{-1}$\,arcsec$^{-2}$
implying a sensitivity of \\
1.0$\times$10$^{-17}$\,erg\,s$^{-1}$\,cm$^{-2}${\AA}$^{-1}$\,arcsec$^{-2}$
at SNR=5.

\section{Data analysis}
\label{anl}

In this section we describe the stellar continuum modeling and the
measurement of the emission line fluxes. The emission-line fluxes are
used to estimate: i) the gas kinematics (Sect.~\ref{GSK}); and in each
galaxy region ii) the line diagnostics (Sect.~\ref{PPG}); iii) the
dust attenuation for the ionized gas and SFR across the galaxy
(Sects.~\ref{SF} and \ref{csf}). The stellar continuum modeling is
needed to obtain the pure emission-line spectrum and, accounting for
the dust extinction, allow us to infer stellar population ages in
different galaxy regions (Sect.~\ref{csf}).

\subsection{Stellar continuum modeling and subtraction}
\label{anl2}

Late-type galaxies present complex star formation histories with
continuous bursts of star formation from the earliest epochs right up
until the present day \citep{K83,KTC94,JPB08,WDJ11}. This implies
that, unlike early-type galaxies, their spectra and absorption line
indices cannot be successfully fit by comparison to those from single
stellar population models. Instead the technique of full spectral
modeling has been developed to reliably describe the properties of the
individual stellar components \citep{CF05,OPL06,KPB09,MGC09}. This
involves the fitting of spectra over extended wavelength ranges
(i.e. not just absorption lines) by linear combinations of multiple
stellar populations, while also accounting for the impact of
line-of-sight stellar motions and instrument resolution, and the
non-linear effects of dust extinction. A variety of implementations
of full spectral fitting have been developed, which explicitly attempt
to derive robust stellar population parameters and their
uncertainties, including the use of detailed simulations
\citep[e.g.][]{CF05,KPO08,KPB09}. 

\begin{figure*}
\includegraphics[width=168mm]{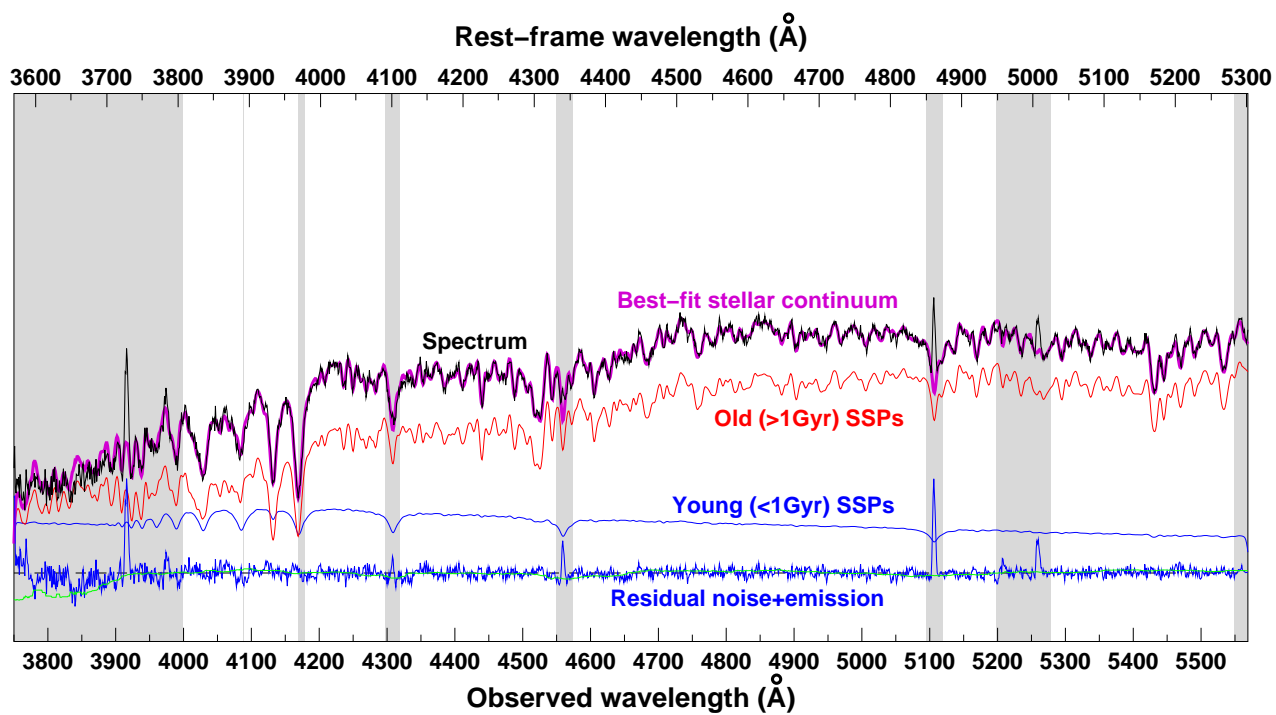}
\caption{Example result from the stellar continuum fitting
process. The black curve shows the input spectrum, coming from the 3x3
spaxels region, centred on the photometric centre of the galaxy. The
magenta curve shows the resultant best-fit stellar continuum
comprising a linear combination of SSPs, requiring both old
($>1$\,Gyr; thin red curve) and young ($<1$\,Gyr old; thin blue curve)
components. The shaded regions indicate the wavelength ranges excluded
from the fitting process, including the masks for emission lines. The
residual emission component (thick blue curve) reveals clear emission
at [O{\sc ii}]\,$\lambda 3729$, H$\delta$, H$\gamma$, H$\beta$ and
[O{\sc iii}]\,$\lambda \lambda\,4959\,5007$.}
\label{fspect}
\end{figure*}

To obtain a reasonable fit to the stellar continuum a signal-to-noise
level of ${\gtrsim}5$0/{\AA} is generally required (e.g. S\'anchez et
al. \citeyear{SRK11}). We take for each spaxel the continuum from the
$3{\times}3$ region centred on the spaxel, which for the main body of
the galaxy was sufficient to achieve the necessary
signal-to-noise. Outside the galaxy disc, further smoothing was
required, and so increasingly large rectangular regions centred on the
spaxel were considered until a mean signal-to-noise level of
${\sim}40$/{\AA} was reached for the stellar continuum over the
wavelength range 4600--4800\,{\AA}.

For each spaxel, the spatially-smoothed spectrum from the blue arm is
firstly fit with the \citet{VSF10} stellar population synthesis
models. We mask out regions below 3950\,{\AA} which have significantly
reduced signal-to-noise levels and flux calibration reliability, and
above 5550\,{\AA} where a bright sky line is located. The wavelength
regions affected by emission lines are also masked out.

We consider a total of 40 simple stellar populations (SSP) models
covering the full range of stellar ages (0.06--15\,Gyr) and three
different metallicities [M/H]=--0.41, 0.0, +0.22. The majority of SSP
models had solar metallicity and provided the required fine sampling
of age, while additional evolved (1--15\,Gyr old) super-solar models
were included to improve the fitting of the small-scale continuum
features. The models assume a \citet{K01} initial mass function
(IMF). The \cite{VSF10} SEDs, based on the Medium resolution INT
Library of Empirical Spectra (MILES) of \citet{SBJ06}, have a nominal
resolution of 2.3\,{\AA}, close to our instrumental resolution, and
cover the spectral range 3540--7410\,{\AA}.

\begin{figure*}
\includegraphics[width=168mm]{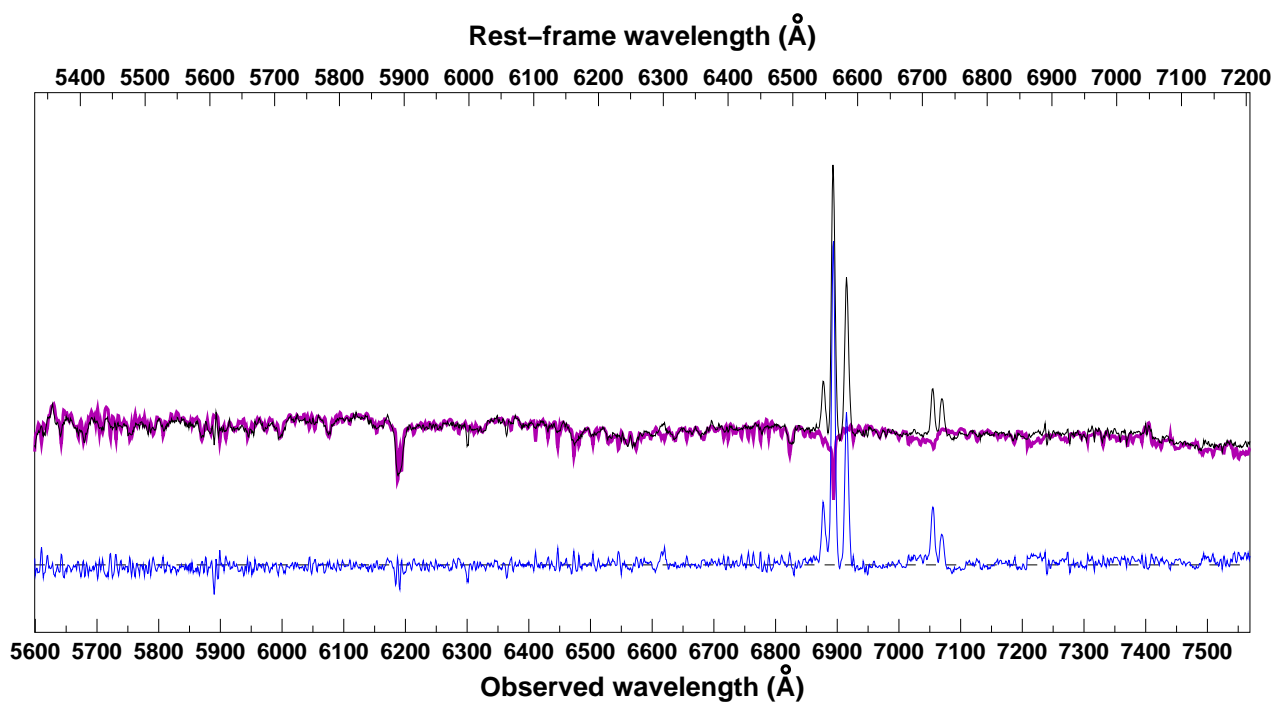}
\caption{Example of the stellar continuum modeling in the red arm. The
black curve shows the red-arm spectrum, coming from the same
3$\times$3 spaxel region as Fig.~\ref{fspect}, which is located in the
core of the galaxy. The magenta curve shows the resultant best-fit
stellar continuum comprising a linear combination of SSPs, based
solely on the fit to the blue arm, and extrapolated to the red
arm. The blue curve shows the residual emission, i.e. subtracting the
magenta curve from the black curve. Notice the emission lines of
[O{\sc i}], H$\alpha$, [N{\sc ii}], [S{\sc ii}] at the observed
wavelengths 6620\AA, 6895\AA, 6880-6917\AA, 7057-7072\AA.}
\label{fig_red}
\end{figure*}

In a first pass, the best-fitting Vazdekis et al. SSP is taken to
derive the absorption-line redshift, $z_{abs}$, and subsequently a
first estimate of the velocity dispersion, $\sigma$, assumed to be in
the range 0--300\,km\,$s^{-1}$. The value of $\sigma$ can be strongly
affected by template mismatch, and so it is important that the full
range of SSP models are fitted over. The $\sigma$ is determined as
that which minimizes the $\chi^{2}$ value for all combinations of
$\sigma$ and SSP model. We then fix the values of $z_{abs}$ and
$\sigma$, and iteratively fit the observed spectrum by a non-negative
linear combination of the 40 SSPs in order to minimize the $\chi^{2}$
value, until convergence is reached. Throughout this iterative
process, individual strongly discrepant pixels (${>}5{\sigma}$) are
identified and masked if necessary, to ensure that single pixels do
not bias the convergence of the fit. This final linear combination is
typically found to be composed of 8--10 individual stellar
populations. A final fit to estimate $\sigma$ is made using the
best-fit complex stellar population model. The final linear
combination of SSPs is then re-scaled linearly to obtain the best-fit
to the spectrum from the single central spaxel and then subtracted to
obtain the pure emission spectrum for that spaxel corrected for
stellar absorption. We neglect internal extinction in this continuum
fitting process, since our aim here is to fit the stellar continuum in
order to model and subtract it in the regions of the emission lines,
while we include dust extinction in our later star formation history
analysis limited to three galaxy regions ({Sect.}~\ref{csf}).

Following \citet{MKT10}, any remaining residuals (typically of the
order a few percent), due to imperfect sky subtraction, or template
mismatch, are removed using a 25-pixel sliding median (again masking
out the emission lines). This last stage had little impact on the
residual spectrum, except for the extreme blue end with
$\lambda{<}3800$\,{\AA}, where the flux calibration accuracy is
weakest, and the additional masking of emission lines, ensured that it
did not impact the final emission line measurements.

Having fit the stellar continuum for the blue arm spectrum, this
best-fit linear combination of SSPs is taken and extended into the red
arm. Then varying only the global scaling factor to account for any
(slight) mismatch in the flux calibration between red and blue arms,
it is subtracted from the red arm spectrum, to produce the pure
emission spectrum for the corresponding spaxel in the red arm.

In Fig.~\ref{fspect} we show an example output of our stellar
continuum fitting process. The input spectrum (black curve) is from a
3$\times$3 spaxels region in the centre of the galaxy. The resultant
best-fit stellar continuum (magenta curve) comprising a linear
combination of SSPs, requires both young and old (thin blue/red
curves) components. Almost all of the small-scale structures seen in
the observed spectrum are real rather than noise, being precisely
mapped by the model stellar continuum. The shaded regions correspond
to wavelength ranges not considered in the fitting process, including
the masks for emission lines. This stellar continuum is subtracted
from the observed spectrum to produce the residual emission component
(thick blue curve) accounting now for stellar absorption, revealing
clear emission at O{\sc ii}, H$\delta$, H$\gamma$, H$\beta$ and [O{\sc
iii}]. The Balmer emission lines are all located within deep
absorption features, demonstrating the necessity of accurately
modeling and subtracting the stellar continuum prior to measuring
these lines. Outside of the emission lines, the rms levels in the
residual signal are consistent with expectations from photon noise,
with little remaining structure. This holds true throughout the galaxy
indicating that on a spaxel-by-spaxel basis the model fits to the
stellar continuum are formally good
($\chi^{2}_{\nu}{\sim}1$). Figure~\ref{fig_red} shows the fit extended
in the red arm for the same spatial region as Fig.~\ref{fspect}. The
SSP model continuum (thick magenta curve) of the spectrum (black
curve) is able to describe all the absorption features and once
subtracted (blue curve) allows robust emission line measurements.

\subsection{Emission-line measurements}
\label{anl1}

Emission-line fluxes and widths were measured with a purposely written
FORTRAN code, which performs a Gaussian fit to the emission lines.
Where lines are either partially overlapping or close (as for the
groups [N{\sc ii}]-H$\alpha$-[N{\sc ii}], [S{\sc ii}]6717-6731 and
[O{\sc iii}]4959-5007), the lines are fitted simultaneously. The fits
are however left completely independent even in the case when their
flux ratios are known to be fixed. This was done to assess the
reliability of our measurements (see below). The code first fits a
straight line to the residual continuum (after sky subtraction and
stellar population fitting and subtraction) on either side of the
line, and then fits a Gaussian to each continuum-subtracted line to
derive the position of the maximum, the peak amplitude, and dispersion
$\sigma$. These values are then used to perform a series of (50)
simulations in which random scatter derived from the continuum fit is
added to the solution, and these simulated lines are fitted again. The
standard deviation of the distribution of the fitting parameters is
then adopted as the contribution of the fit to the uncertainty. Fluxes
are computed from the measured amplitudes and $\sigma$.

We estimate the SNR of the emission as being the ratio of the peak
amplitude to the standard deviation of the surrounding
continuum. Comparing the SNR with the error coming from the fit and
taking into account the appearance of the lines, we fix at SNR=5 the
lower limit for reliable measurements. This limit corresponds to
$\sim$30\% relative errors from the fit for both the flux and
$\sigma$. The total error on the flux is computed adding in
quadrature the uncertainty in the flux zero-point, estimated comparing
the fluxes of the individual exposures taken in good atmospheric
conditions. This amounts to $\sim$7\% --10\% ($R$ and $B$ respectively).

\begin{figure}
\includegraphics[width=80mm,height=170mm]{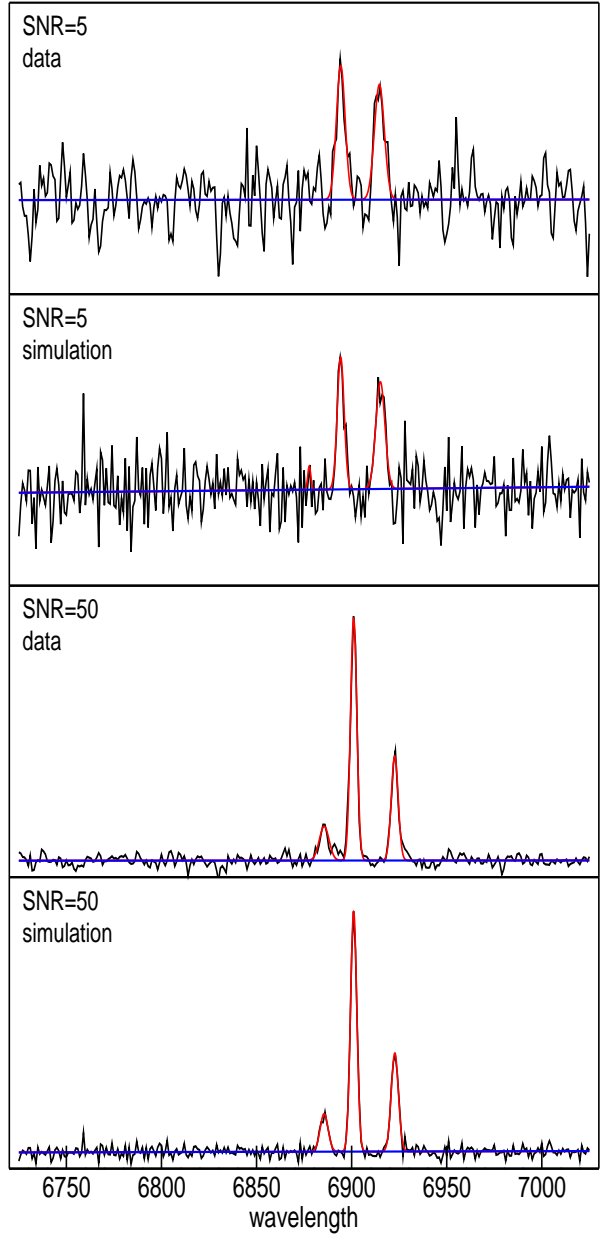}
\caption{Measurement of the emission lines in the group [N{\sc
    ii}]-H$\alpha$-[N{\sc ii}] for two values of SNR of
  H$\alpha$. From top to bottom the real data and one of the
  simulation in a spaxel with SNR$\sim5$ (first two plots) and in a
  spaxel with SNR$\sim50$ (last two plots). The black curves are the
  observed and simulated spectra, while the red curves are the fits to
  the emission lines and the continuous lines are the fits to the
  continuum. SNR=50 is the mean value of our data.}
\label{hafit}
\end{figure}

Figure~\ref{hafit} shows examples of both real and simulated data, and
the computed fits in the case of SNR$\sim$5 (our lower limit) and
SNR$\sim$50 (the average of our data) in the wavelength region around
H${\alpha}$.

The instrumental contribution to the line widths was estimated by
measuring the widths of selected arc-lamp lines from the $B$ and $R$
data-cubes. Due to the combination of spherical aberration and
high-order astigmatism in the instrumental point spread function, the
instrumental line width $\sigma_{instr}$ depends on the wavelength and
on the X spatial direction, but not on the Y direction. We constructed
maps of $\sigma_{instr}(X,\lambda)$ which were subtracted in
quadrature from the data to obtain the intrinsic velocity dispersion
of the gas. At our observed wavelength of H$\alpha$
(${\sim}6$800\,\AA), $\sigma_{instr}{\sim}5$1--62\,km\,s$^{-1}$, being
close to the maximum in the centre of the field. The total error on
the velocity dispersion accounts for the uncertainty of
${\sim}5$\,km\,s$^{-1}$ in $\sigma_{instr}$. The errors from the fit
on radial velocity are very small (${<}5$\,km\,s$^{-1}$), so the main
contribution comes from the uncertainty in wavelength calibration,
which is ${\sim}13$\,km\,s$^{-1}$.

As explained above, we chose to measure all lines independently in
order to be able to assess the reliability of our flux ratios. With
this aim, we compared our measurements of the [N{\sc ii}]
$\lambda\lambda$6583/6548 and [O{\sc iii}] $\lambda\lambda$4959/5007
line ratios to their theoretical values of 2.976 \citep{DS03} and
2.936 \citep{SZ00}, respectively. We found that for an
error on the flux ratio less than 30\%, the involved lines must have
at least SNR$>$10 (consistent with the above errors on the individual
fluxes), and adopted a more conservative limit of SNR$>$20.

To achieve this SNR, we applied a spatial adaptive binning to our data
by means of the \emph{`Weighted Voronoi Tessellation'} (WVT) by Diehl
and Statler \footnote{http://www.phy.ohiou.edu/diehl/WVT}
(\citeyear{DS06}), which is a generalization of the algorithm that
\citet{CC03} developed for SAURON data. The WVT performs the
partitioning of a region based on a set of points called `generators',
the points around which the partition of the plane takes place. The
partitioning is repeated iteratively until some target SNR is achieved
in all bins. The WVT algorithm allows to `manually' set a number of
generators, which we chose on the basis of the position with respect
to the stellar disc. We obtained a total of 104 regions. As input data
we used the signal and noise of the H$\alpha$ emission line, which has
the widest spatial coverage. The faintest line involved in our flux
ratios is the [O{\sc i}]\,$\lambda$6300 line, whose flux is typically
about $\sim20$\% of H$\alpha$, so that we set SNR=100 for H$\alpha$ as
the target SNR for the WVT algorithm in order to achieve the required
SNR for the [O{\sc i}] line. The WVT is adopted to derive flux ratios,
dust attenuation, and SFR, but not for the kinematics, for which no
spatial binning was necessary.

\begin{figure}
\includegraphics[width=84mm]{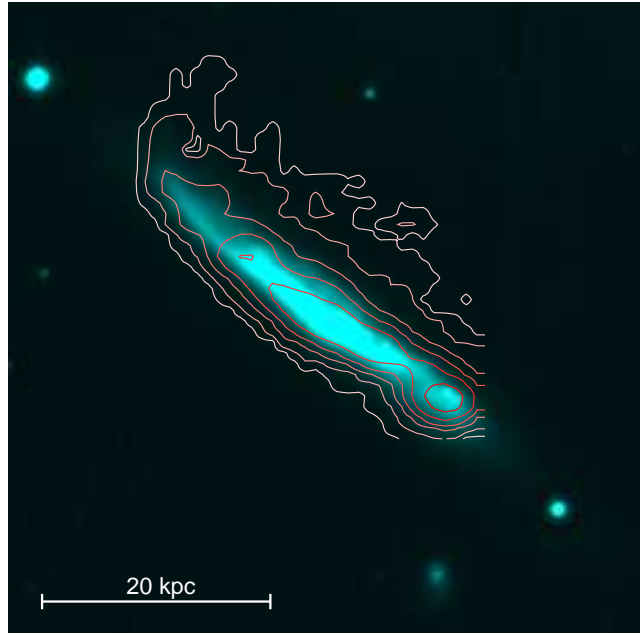}
\caption{Logarithmically-spaced contours of H$\alpha$ flux in the
  range 5$\cdot$10$^{-18}$--
  5$\cdot$10$^{-16}$erg\,cm$^{-2}$s$^{-1}$\AA$^{-1}$arcsec$^{-2}$ from
  WiFeS are superimposed on the ESO-WFI $R$-band image of the
  galaxy. Extraplanar gas is visible out of the disc in the NW (upper
  right). The scale is shown on the bottom. The H$\alpha$ flux was
  corrected for dust attenuation (Sect.~\ref{SF}).}
\label{foutflow}
\end{figure}

\section{Results}
\label{res}

Figure~\ref{foutflow} shows the contours of the H$\alpha$
emission-line flux from WiFeS tracing the ionized gas superimposed on ESO-WFI
$R$-band image. The ionized gas extends for about 13\,kpc in
projection out of the galaxy disc in the NW direction. No gas is
detected beyond that distance, although pointing \#2 covers further
13\,kpc North from the most external gas isophote in
Fig.~\ref{foutflow}. In order to understand the origin of the
extraplanar gas, we examine the structure and morphology of the
ionized gas, derived the gas and stellar kinematics and the dust
extinction across the galaxy.

\subsection{Morphology of the H$\alpha$ emission} 
\label{Ha_morph}

The H$\alpha$ narrow-band image allows us to resolve the structure of
the extraplanar gas. Fig.~\ref{halpha} shows the MMTF H$\alpha$ image
(white) combined with the UKIRT $K$-band (red) image (left panel) and
with the contours of H$\alpha$ flux measured with WiFeS (right
panel). The most noticeable feature of the H$\alpha$ image are the
compact (${\lesssim}0.5^{\prime\prime}$; ${\lesssim}500$\,pc) knots of
H$\alpha$ emission seen all along the NW side of the disc. Some of
these knots seem ``tethered'' to the disc by faint filamentary
strands. The most distant knot is seen at $\sim$13\,kpc (in
projection) from the galaxy major axis, but most knots are much
closer, within 3--4\,kpc of the disc. All of these knots and
filamentary structures are completely absent in the MMTF continuum
image, confirming that this is from H$\alpha$ emission with little if
any underlying stellar continuum component. It is notable that there
are no such H-alpha features on the SE side.

\begin{figure*}
\includegraphics[width=160mm]{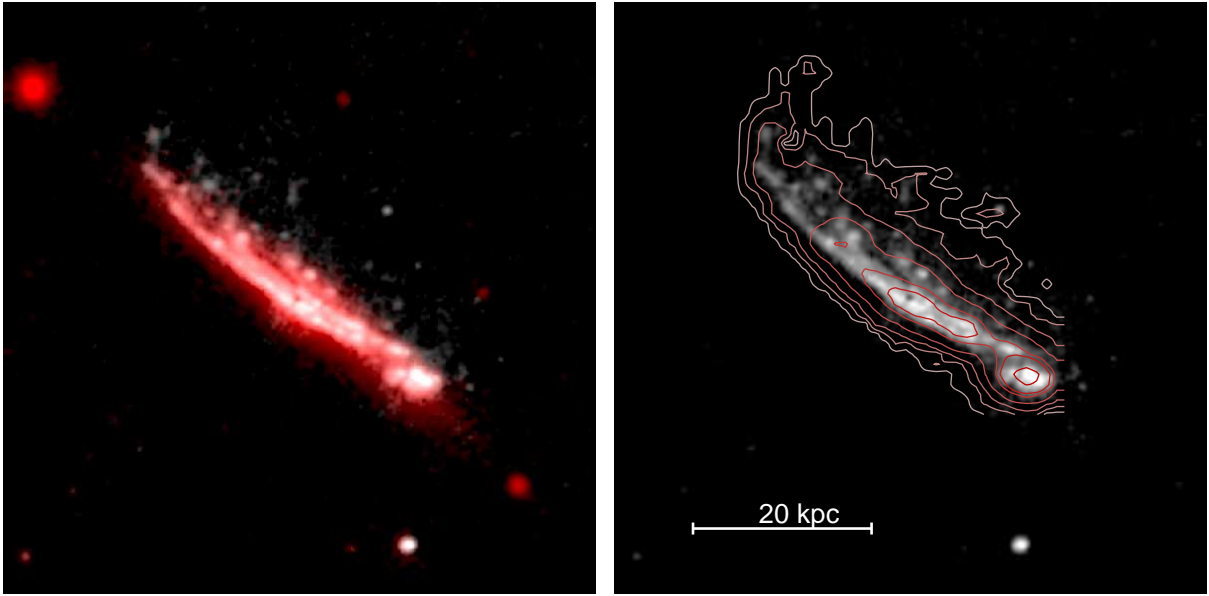}
\caption{{\it Left}: Composite image of SOS\,114372 consisting of the
  MMTF H$\alpha$ image ({\em white}) over the UKIRT $K$-band image
  ({\em red}). SOS\,115228 is seen as the white spot closest to the
  bottom. The H$\alpha$ flux is not corrected for dust
  attenuation. {\it Right}: Contours of H$\alpha$ flux measured with
  WiFeS over the MMTF H$\alpha$ image. The scale for the two panels is
  indicated.}
\label{halpha}  
\end{figure*}

The most luminous knot has an H$\alpha$ luminosity of
$7.2{\times}10^{38}$\,erg\,s$^{-1}$, while the faintest detected knots
have $L(H{\alpha}){\sim}1.5{\times}10^{38}$\,erg\,s$^{-1}$. If all of
this emission were powered by star formation, these luminosities would
correspond to SFRs of 0.0008--0.004\,M$_{\odot}$\,yr$^{-1}$ based on
the calibration of \citet{K98} with a Kroupa IMF (see
Sect.~\ref{PPG}). A bright, clumpy H$\alpha$-emitting region
${\sim}$6\,kpc in extent is located in the disc $\sim$12\,kpc SW of
the galaxy centre, indicative of a localized ongoing starburst (see
Sect.~\ref{SF}).

The small galaxy SOS\,115228 (Fig~\ref{SOS}, right) SE of SOS\,114372
has also an H$\alpha$-emitting region associated with it, which
appears ${\sim}2.8{\times}$ brighter in our MMTF H$\alpha$ filter than
the continuum filter. Although SOS\,115228 does not have a known
redshift, this might be suggestive of a starbursting dwarf galaxy at
the same redshift (within ${\sim}400$\,km\,s$^{-1}$) as \\
SOS\,114372. Although clearly compact, SOS\,115228 does not appear as
a point source in H$\alpha$, but likely has an intrinsic diameter of
${\sim}0.35^{\prime\prime}$ (350\,pc). Alternatively, SOS\,115228
could be a much more distant active (starburst or AGN) galaxy whose
H$\beta$ emission happens to be detected in our MMTF image. The
redshift of this galaxy would then be z$\sim$0.42. This difference in
redshift would be consistent with the $\sim$5\,mag difference in $K$-band
magnitude with SOS\,114372, if the two galaxies have comparable
masses. Unfortunately, no conclusion can be drawn on the nature of
this object without spectroscopic data.

\subsection{Gas and stellar kinematics}
\label{GSK}

The gas velocity field derived from IFS observations is shown in
Fig.~\ref{fGVF} (left panel). The white contours trace the $R$-band
stellar continuum derived from the spectrum once the emission lines
have been removed. The dashed lines mark the major and minor axes of
the disc extending to 3$\times r_d$, and cross at the $K$-band
photometric centre. The black curves are iso-velocity contours. The
kinematic centre is assumed to coincide with the photometric centre of
the $K$-band image. It turns out that this choice also produces the
smallest asymmetries in the major axis kinematic profile.

\begin{figure*}
\includegraphics[width=145mm]{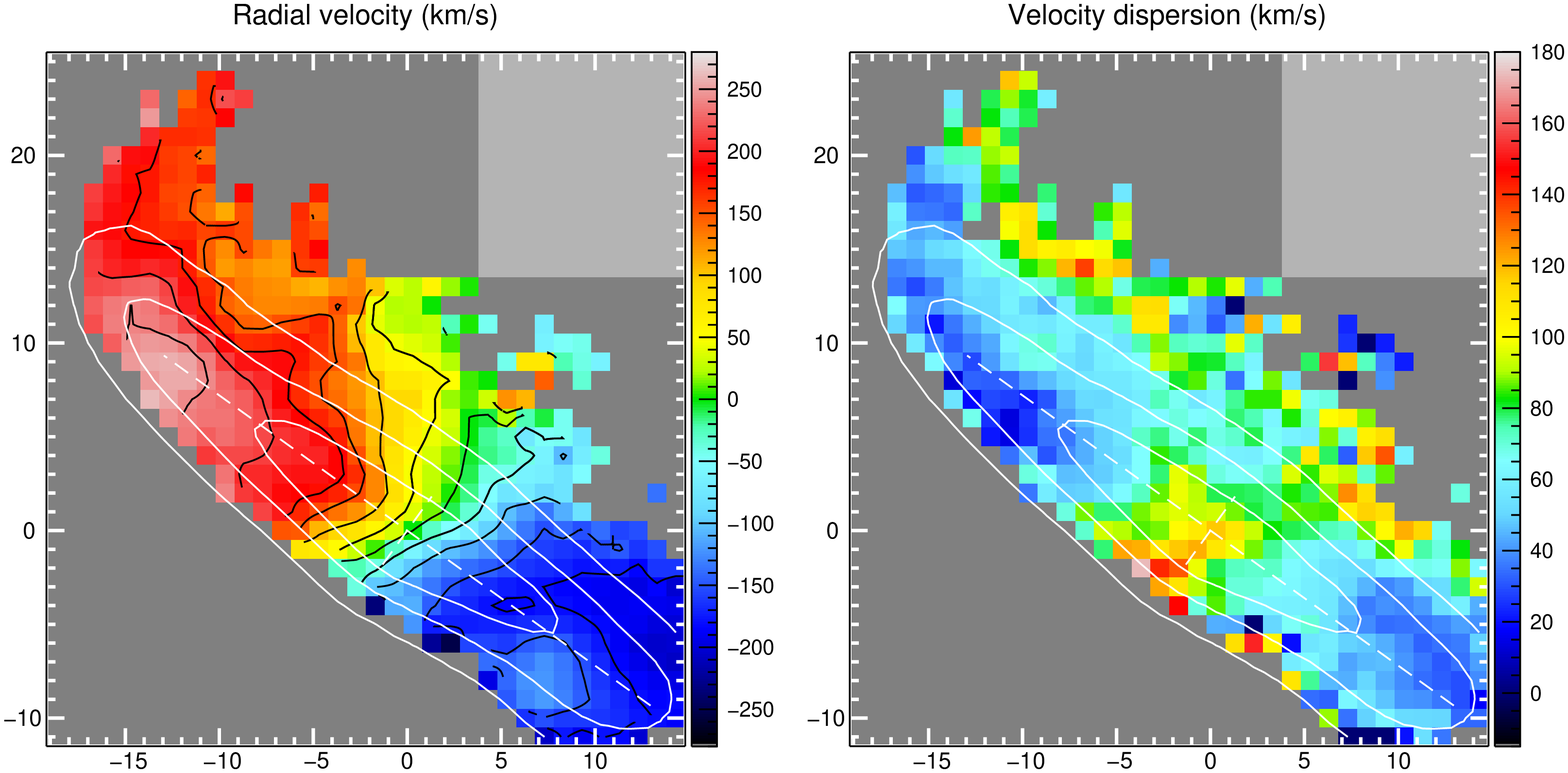}
\caption{Gas kinematics of SOS\,114372 derived from the fit to the
  H$\alpha$ emission line. The light gray rectangle marks the region
  not covered by the two pointings. The white iso-density contours
  trace the galaxy red continuum and the dashed lines, extending to
  3$\times r_d$, show the positions of the minor and major axis and
  cross at the $K$-band photometric centre. {\it Left}: Gas velocity
  field. Each pixel is colour coded according to the measured radial
  velocity relative to the galaxy centre. The black curves are
  iso-velocity contours. {\it Right}: Gas velocity dispersion. Each
  pixel is colour coded according to the measured velocity
  dispersion. In this figure and in the following maps the ticks on
  the axes indicate the apparent distance from the $K$-band
  photometric centre in arcsecs. The velocity scales are on the right
  of each panel.}
\label{fGVF}
\end{figure*}

\begin{figure*}
\includegraphics[width=145mm]{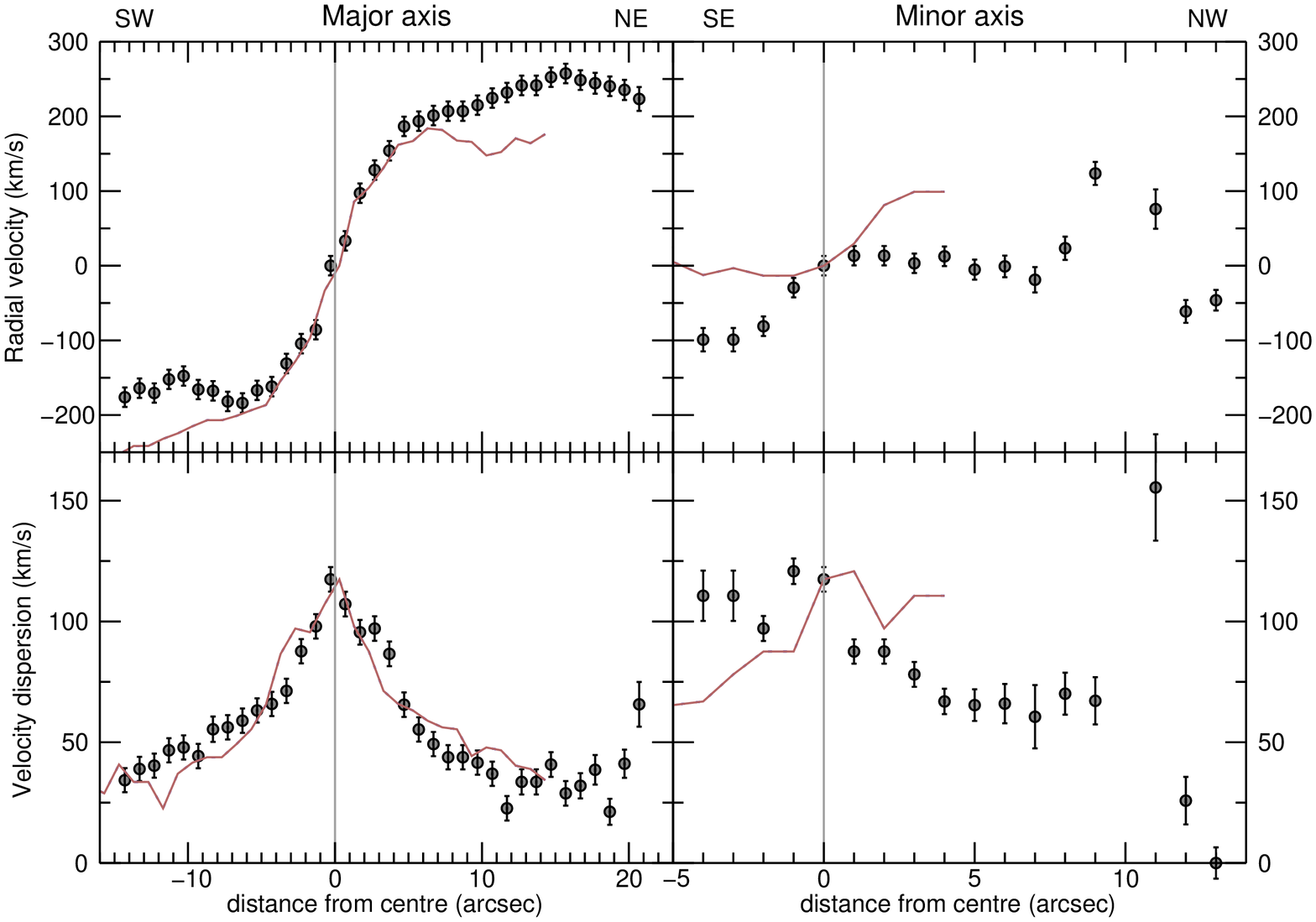}
\caption{Gas line-of-sight velocity (upper panels) and velocity
  dispersion (lower panels) profiles along the major (left panels) and
  minor axis (right panels). The red lines represent the same data
  reflected with respect to the origin, to facilitate the inspection
  of asymmetries. Distances from the photometric centre are positive
  towards NE (major axis) and NW for the minor axis. Directions in the
  sky are given on the top of the figure.}
\label{fGRC}
\end{figure*}

The velocity field is complex. Within the disc, the overall appearance
of the field is that of a rotating disc, although significant
departures from a simple rotation are present everywhere, as it is
also clearly shown in the radial velocity profiles along the major and
minor axes (Fig.~\ref{fGRC}). 

The major axis velocity profile (Fig.~\ref{fGRC}) is fairly
symmetric in the inner $\sim$5\,arsec. Beyond that radius, the radial
velocity in the NE side of the disc increases until it reaches
$\sim$250\,km\,s$^{-1}$ at $\sim$16\,arcsec from the centre and then
it starts decreasing until the last observed point at 21\,arcsec. In
the SW side, the absolute value of the velocity first decreases to a
local minimum of $\sim$140\,km\,s$^{-1}$ at 10\,arcsec from the centre
and then slightly increases until the last observed point. The whole
situation depicted in Fig.~\ref{fGVF} is even more complex than this,
as is immediately clear from the shape of the iso-velocity
contours. The velocity field in the disc is asymmetric also with
respect to the minor axis, with the SE side of the disc generally more
red-shifted than the NW side. There are two local maxima in the SE
side, one at $\sim$3\,$r_d$ NE from the centre and one at
$\sim$2\,$r_d$ SW from it. The kinematics of the extraplanar gas
appears to be remarkably dominated by the rotation characterising the
disc, maintaining the iso-velocity contours continuous up to the most
external limits. We notice that, NE from the galaxy centre, the
extraplanar gas is in the average blue-shifted with respect to the
nearest gas in the disc, while in the SW it is red-shifted, thus
producing 'fan-shaped' iso-velocity contours. The minor axis radial
velocity profile allows us to identify a gas stream approaching the
observer, reaching a velocity of $\sim$100\,km\,s$^{-1}$ at
$\sim$3-4\,arcsec SE from the nucleus.

From the major axis radial velocity profile we derive a dynamical mass
within 20\,kpc of $\mathcal{M}_{dyn}{=}2.1{\cdot}10^{11}$M$_\odot$.

The gas velocity dispersion field is also quite complex
(Fig.~\ref{fGVF}; right panel). We first notice that the velocity
dispersion $\sigma$ rarely reaches values as small as would be
expected in normal turbulent \HII~regions
($\sigma{\sim}2$0--30\,km\,s$^{-1}$). The velocity dispersion is
particularly high ($>100$\,km\,s$^{-1}$) over a significant fraction
of the extraplanar gas. High values of $\sigma$ are also observed in
two regions near the SE border of the disc.  The most remarkable of the
two is the triangular area with $\sigma \gtrsim$120\,km\,s$^{-1}$
extending SE from the nucleus, where we also measure the absolute
maximum of the velocity dispersion (${\sim}$180\,km\,s$^{-1}$). This
area corresponds to the stream directed towards the observer noted
above. We remark that the value of the velocity dispersion also
depends on the spatial sampling of the spectrograph, increasing with
pixel size and seeing, as the combination of the motions of more gas
elements is measured in each resolution element. This is certainly a
major origin of our average high values of $\sigma$, because it is
clear from the radial velocity field that complex motions are taking
place giving rise to different possible kinds of superimpositions. We
notice that our resolution element of
1$^{\prime\prime}\times1^{\prime\prime}$ corresponds to
$\sim$1$\times$7\,kpc$^2$ once projected on the plane of the disc. We
also remark that the presence of large quantities of obscuring dust
may complicate the interpretation of the gas kinematics, since dust
may selectively hide parts of the moving gas (e.g. Giovanelli \&
Haynes \citeyear{GH02}; Baes et al. \citeyear{BDD03}; Kregel et
al. \citeyear{KVF04}; Valotto \& Giovanelli \citeyear{VG04}), although
this would probably not increase the measured velocity dispersion.

\begin{figure}
\includegraphics[width=84mm]{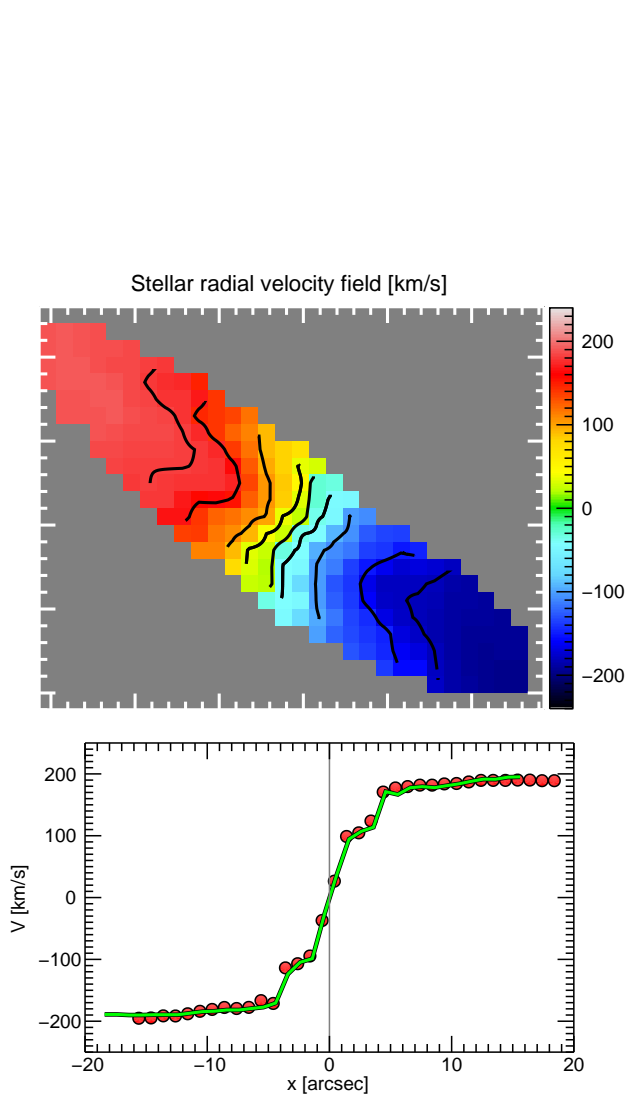}
\caption{{\it Top}: The stellar velocity field with
  Iso-velocity contours. {\it Bottom}: The stellar radial velocity
  profile along the major axis (red dots) with its reflection with
  respect to the centre superimposed (green curve).}
\label{fGSK}
\end{figure}

From the fit of the stellar continuum we derived the velocity field of
the stars. In order to estimate the uncertainties in the stellar
velocity we used simulations in which random noise is added to the
spectra. A typical uncertainty of $\sim$2\,km\,s$^{-1}$ for spaxels in
the disc is found. This low uncertainty is probably due to the
previous 3$\times$3 smoothing in the spatial direction. Including the
uncertainty in wavelength calibration, this sums up to
$\sim$14\,km\,s$^{-1}$.

The derived stellar radial velocity field and major axis radial
velocity profile are shown in Fig.~\ref{fGSK}. The iso-velocity
contours of the stellar component appear symmetric with respect to the
centre, while they are skewed with respect to the minor and major
axis. Along the major axis, the radial velocity grows linearly and
rapidly in the inner $\sim$1.5-2\,arcsec, where there is an abrupt
change of slope followed by another rapid increase. At $\sim$5\,arcsec
from the centre the radial velocity assumes an almost constant value
of $\sim$180\,km\,s$^{-1}$ until the last observed points. The
kinematic centre of the stellar velocity field is displaced
0.3\,arcsec (280 pc) NE with respect to the photometric ($K$ band)
centre of the galaxy (and consequently to the kinematic centre of the
gas). However, we will not further discuss this point, since this
displacement is smaller than our resolution element.

\begin{figure}
\includegraphics[width=84mm]{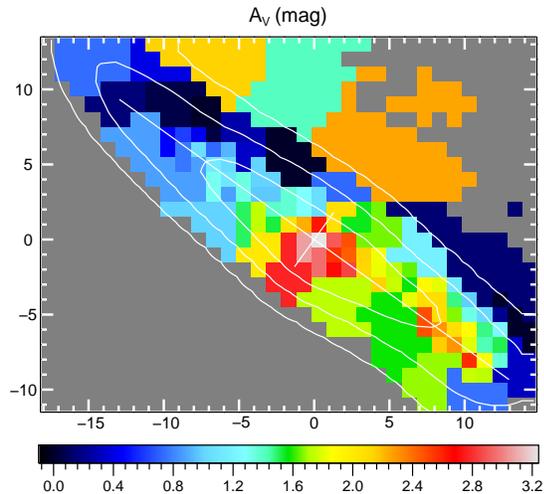}
\caption{Dust attenuation derived from the H$\alpha$/H$\beta$ line
ratio.}
\label{fatt}
\end{figure}

\subsection{Dust extinction across the galaxy}
\label{SF}

Figure~\ref{fatt} shows the distribution of the dust attenuation
derived from the H$\alpha$/H$\beta$ line ratio, in terms of the visual
extinction $A_V$. We use the theoretical attenuation curve by
\citet{FD05} with $R_{V}{=}4.5$. The dust extinction is estimated in
each of the binned region of Fig.~\ref{WVT}. The level of dust
attenuation is notably asymmetric along the major axis, being
universally lower on the NE side of the disc than the SW side. The
highest dust extinction ($A_V{\sim}2$.4--3.2\,mag) is observed in the
centre of the galaxy, extending along the minor axis to the SE
edge. Highly attenuated regions ($A_V{\sim}1$.9--2.6\,mag) are also
found in the SW disc ${\sim}1$2\,kpc from the galaxy centre and in the
extraplanar gas. This extinction map agrees with what can be inferred
from the optical images, where dust absorption is clear in the galaxy
centre (see also Fig.~\ref{SOS} right panel).

\section{Properties of the gas and star formation across the galaxy}
\label{gas_sf}

Our data, complemented by available multi-band imaging (from far-UV to
far-IR), allow us to investigate the nature of the gas and the star
formation across the galaxy, but first we examine the gas and stellar
kinematics.

The stellar kinematics is typical of a barred galaxy, with the
twisting of the iso-velocity contours due to the non-circular motions
induced by the bar \citep{AM02} and with a `double-bump' rotation
curve \citep{ChB04, BuA05}. A detailed analysis of the stellar
kinematics is beyond our goals, since it would require much higher
spatial resolution (e.g. Emsellem et al. \citeyear{ECK07}), but the
important point for this work is that the stellar velocity field
appears to be regular and symmetric, without any signs of
perturbation. One may ask whether the regularity of the stellar
kinematics field, in sharp contrast with the gas, is due to the
smoothing applied to the spectra to increase the SNR of the stellar
component. We verified that this is not the case by applying the same
3$\times$3 binning to the gas velocity field. The effect was just a
smoothing of the small-scale features, with the gross properties
remaining unchanged. This holds also for much more severe binning (up
to 7$\times$7 pixels). The complexity of the gas velocity field cannot
be explained by the presence of the bar alone, as can be seen by
comparing the iso-velocity contours in Fig.~\ref{fGVF} to the velocity
fields of barred galaxies reported, for instance, in \citet{B81},
\citet{CAB03}, \citet{FVP05} and \citet{DNR09}. All these
considerations make us to conclude that the gas kinematics is heavily
perturbed, while the stellar kinematics is not.

\subsection{Physical properties of the gas}
\label{PPG}

The key diagnostics for the examination of the mode of excitation of
an ionized plasma were introduced by \citet{VO87}. These use the
[N{\sc ii}]/ \Ha, [S{\sc ii}]/\Ha or [O{\sc i}]/\Ha ratios plotted
against the [O{\sc iii}]/\Hb ratio, and have the great advantage of
being affected very little by dust extinction. They allow a detailed
classification of the excitation as either due to young stars, or due
to an active nucleus. The classification scheme has been refined by
Kewley et al. (\citeyear{Kewley01}), Kauffmann et
al. (\citeyear{Kauffmann03}) and Kewley et al. (\citeyear{Kewley06})
by the use of more sophisticated models to define the allowed range of
ratios produced in \HII~regions, and using the SDSS galaxy sample to
create a semi-empirical classification to distinguish between Seyferts
or LINER-like AGN.

\begin{figure}
\includegraphics[width=84mm]{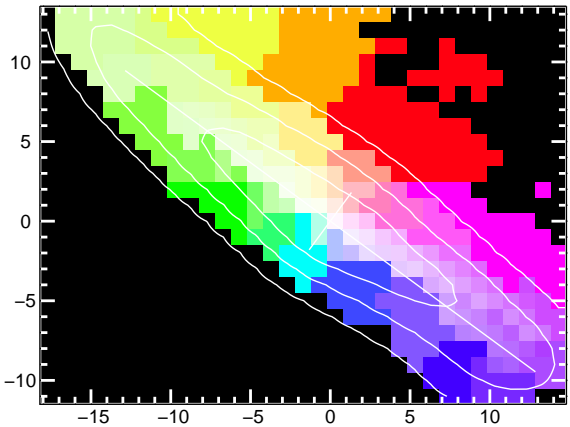}
\caption{The 104 galaxy regions identified by the WVT are shown in
different colours. For each of these regions reliable and robust
measurements of emission line ratios are derived. The colours used in
this figure are consistently used in Figs.~\ref{fbpt3} and
\ref{fvsk}.}
\label{WVT}
\end{figure}

We investigate the physical properties of the gas using the pointing
\#1 where the suitable SNR (SNR=100 for H$\alpha$, see
Sect.~\ref{anl1}) is achieved by binning the data through the WVT. In
Fig.~\ref{WVT} we show the 104 different galaxy regions where
independent measurements of the spectral indices were derived. These
include regions outside of the disc where the extraplanar gas is detected
although with a lower SNR per spaxel.

\begin{figure*}
\includegraphics[width=160mm]{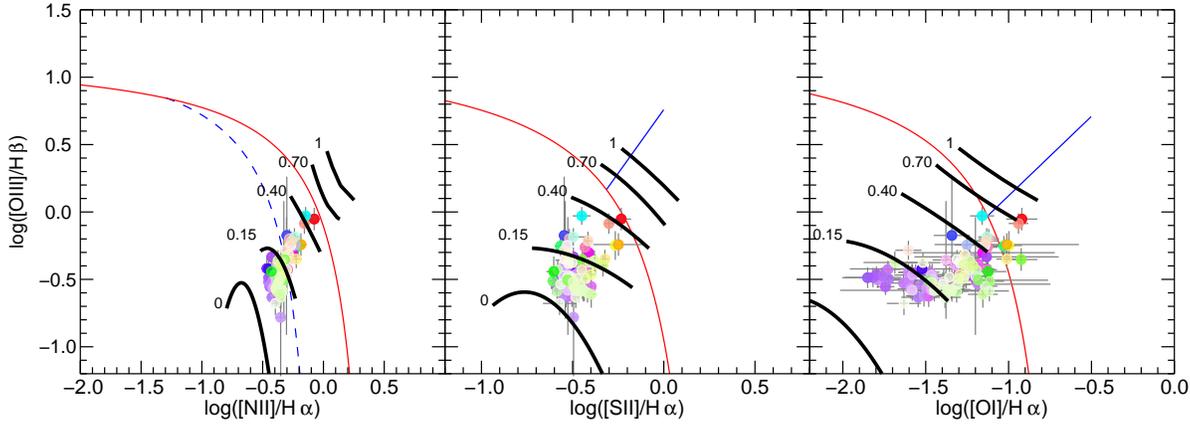}
\caption{Line flux diagnostic diagrams of the different regions of
 SOS\,114372 (see text). The shock and photoionization models by
 \citet{RKD11} are superimposed on the measured flux ratios. The black
 curves are drawn for different fractions (from 0 to 1) of H$\alpha$
 flux contributed by shocks as indicated. For comparison the
 theoretical (red curve) and empirical (blue dashed curve in the left
 panel) upper limits for \HII~regions are also indicated as well as
 the separation between AGN and LINER (solid blue line in the central
 and right panel).}
\label{fbpt3}
\end{figure*}

In Fig.~\ref{fbpt3} we show the measured line ratios for the different
galaxy regions, colour coded as in Fig.~\ref{WVT}, along with the
theoretical predictions and limits for photoionization and shock
excited models. The red curves mark the theoretical upper limit for
\HII~regions, while the dashed blue curve in the left panel delineates
the empirical limit to star-forming regions defined by
\citet{Kauffmann03}. Between the red curve and the dashed blue curve,
`composite' systems are found whose emission is characterized by a
mixture of AGN and \HII~regions. The blue continuous line in the
central and right panels divides the line ratios typical of Seyfert
galaxies (upper part of the diagram), from those of low ionization
emission line regions (`LINER'). These curves are extensively
used in the literature and plotted here to facilitate comparison with
other works.

The black curves represent shock and photoionization models from
\citet{RKD11}, in which a fraction of the radiation (as indicated by
the numbers close to the curves) is produced by shock excitation and
the remaining comes from photo-ionization. These models are computed
using the MAPPINGS IIIr code, the latest version of the code
originally introduced in \citet{Sutherland93}. The abundance set is
taken from~\cite{Grevesse10} and we use the standard (solar region)
dust depletion factors from \citet{Kimura03}. The shock models have
velocities in the range 100 to 200\,km\,s$^{-1}$ in steps of
20\,km\,s$^{-1}$, with hydrogen assumed to be fully pre-ionized. In
using these depletion factors we are implicitly assuming that the
shock is not sufficiently fast to sputter the dust grains advected
into the shocked region. The models presented here have
Z/Z$_{\odot}{=}2$ (\OH$=8.99$), and have a transverse magnetic field
consistent with equipartition of thermal and magnetic energy
($B{=}5$\,$\mu$G for $n_c{=}10$\,cm$^{-3}$; $B{\propto}n_c^{-1/2}$).

The photoionization models are generated using UV flux distributions
appropriate to continuous star formation computed using the
Starburst99 code \citep{leitherer99}. We have used the older models
employed by \citet{Kewley01} rather than more recent ones, since the
hardness of the radiation field provides a closer match to the
\HII~region line ratios observed in the SDSS galaxy sample. We have
also adjusted the nitrogen abundance to provide a still closer match
to the SDSS galaxies. The photoionization models are characterized by
an ionization parameter $6.5{<}\log[q]{<}7.75$\,cm\,s$^{-1}$. This
ionization parameter can be converted to the more commonly used
dimensionless ionization parameter $U$ by $U{=}q/c$, where $c$ is the
speed of light.

We have created mixing models between each photoionization model, and
all shock models. These are characterized by a fraction $f$ of the
flux at \Ha being contributed by shocks. These models do not create a
simple mixing line on the diagnostic plots, but rather a zone within
which the line ratios are consistent with a certain mixing fraction of
shock excitation. In Fig.~\ref{fbpt3} we show the average of the line
ratios produced by mixing fractions, $f {=}0$, 0.15, 0.4, 0.7 and
1.0. Note that the line ratios are sensitive to quite small
contributions from shock emission - as little as 15\% contribution by
shocks can change observed line ratios by a factor of two.

The observed line ratios cannot be simply explained by \HII-like
photoionization, but display a spread which is characteristic of a
mixture of both shock excitation and photoionization. All parts of the
galaxy seem to be contaminated by at least a small fraction of shock
excitation, $0.05 {\lesssim} f {\lesssim} 0.1$. The regions which are
dominated by star formation and the associated \HII~regions are
preferentially located in the SW portion of the galaxy (purple
points). Regions in the extraplanar gas to the NW of the galaxy
(red/orange points) have much higher shock excitation fractions,
ranging up to 0.4--0.8. In addition, there is a strong concentration
of shock-excited gas along the minor axis SE from the nucleus (light
blue points) and in another Northern region of the SE edge (green
point).

\subsection{Star formation across the galaxy}
\label{csf}

\begin{figure}
\includegraphics[width=84mm]{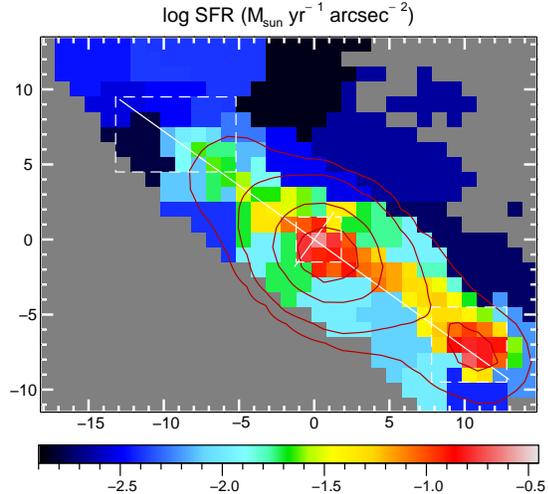}
\caption{The logarithmic SFR density across the galaxy derived from
  H$\alpha$ flux corrected for the extinction and shock-excitation,
  where necessary. The contours show the isophotes of the 24$\mu$m
  flux. The three boxes outlined by white dashed sides mark the
  regions of the galaxy where the stellar population analysis was
  performed. The three regions are, from right-bottom to top-left: the
  starbursting region in the SW disc, the central bulge region, and
  the NE disc.}
\label{fsf24}
\end{figure}

To study the star formation across SOS\,114372, we first analyze the
ongoing star formation from the flux of the H$\alpha$ emission line.
Then we study the recent star formation history from the distribution
of the ages of stellar populations in three distinct regions of the
galaxy - a starbursting region in the SW disc (see below), the
central bulge region, and the NE disc (see Fig.\ref{fsf24}).

\subsubsection{Ongoing star formation}

We derive the current SFR from the H$\alpha$ flux taking into account
the effects of dust extinction following \citet{K98}.  For the
derivation of the SFR we approximately removed the contribution of the
shock-ionized gas to the H$\alpha$ flux -- the H$\alpha$ fluxes of the
galaxy regions with $\ge$50\% of emission coming from shock-ionized
gas are reduced by a factor of two. This correction is however very
small compared to the uncertainties. The logarithm of SFR surface
density across the galaxy is given in Fig.~\ref{fsf24}.  Intense star
formation is found in the disc spanning from 0.01 to
0.34\,M$_\odot$\,yr$^{-1}$arcsec$^{-2}$ with the global maximum
located in the galaxy centre. There is also a notable local maximum in
the SW region of the disc at $\sim$ 12 kpc from the centre, with
SFR=0.2\,M$_\odot$\,yr$^{-1}$arcsec$^{-2}$. The contours in
Fig.~\ref{fsf24} show the distribution of the 24$\mu$m flux which, in
agreement with the SFR derived form H$\alpha$, presents the maximum in
the centre and the second peak in the SW side. These two regions are
also the most obscured regions identified in Fig.~\ref{fatt}.
These results imply a coherent scenario where the current SFR, derived
from the H$\alpha$ emission and the 24$\mu$m flux, occurs in the
highly dust-extinguished molecular clouds. The integrated
H$\alpha$-derived SFR of SOS\,114372 (adopting the Kroupa IMF) amounts
to 7.2$\pm$2.2\,M$_{\odot}$\,yr$^{-1}$. The error takes into account
the uncertainties related to the flux and attenuation measurements
added to a 30\% uncertainty due to different calibrations of H$\alpha$
as a SFR indicator \citep{K98}. From the difference between raw and
attenuation-corrected H$\alpha$ fluxes we can estimate the obscured
SFR as 5.3$\pm$1.6\,M$_\odot$\,yr$^{-1}$. This obscured SFR is also
measured by the IR dust emission turning out to be
6.65$^{+1.99}_{-1.19}$\,M$_\odot$\,yr$^{-1}$. The two independent
estimates of the obscured SFR are therefore fully consistent.

\begin{figure*}
\includegraphics[width=150mm]{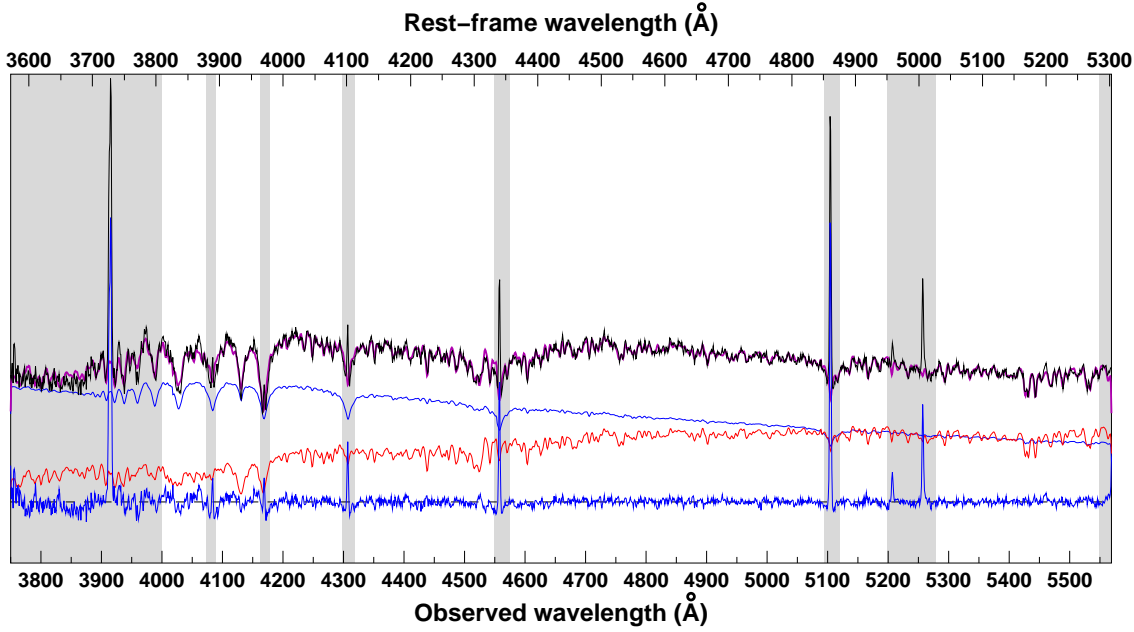}
\caption{The integrated spectrum (black curve) and full population
synthesis fit (magenta) for the starburst region in the SW side of the
disc. The contributions to the model fit from the old (${>}1$\,Gyr;
thin red curve) and young (${\le}1$\,Gyr; thin blue) stellar
populations are shown. The shaded regions indicate the wavelength
ranges excluded from the fitting process. The residual emission
component after subtraction of the model stellar continuum is shown by
the thick blue curve.}
\label{ssplower}
\end{figure*}

\begin{figure*}
\includegraphics[width=150mm]{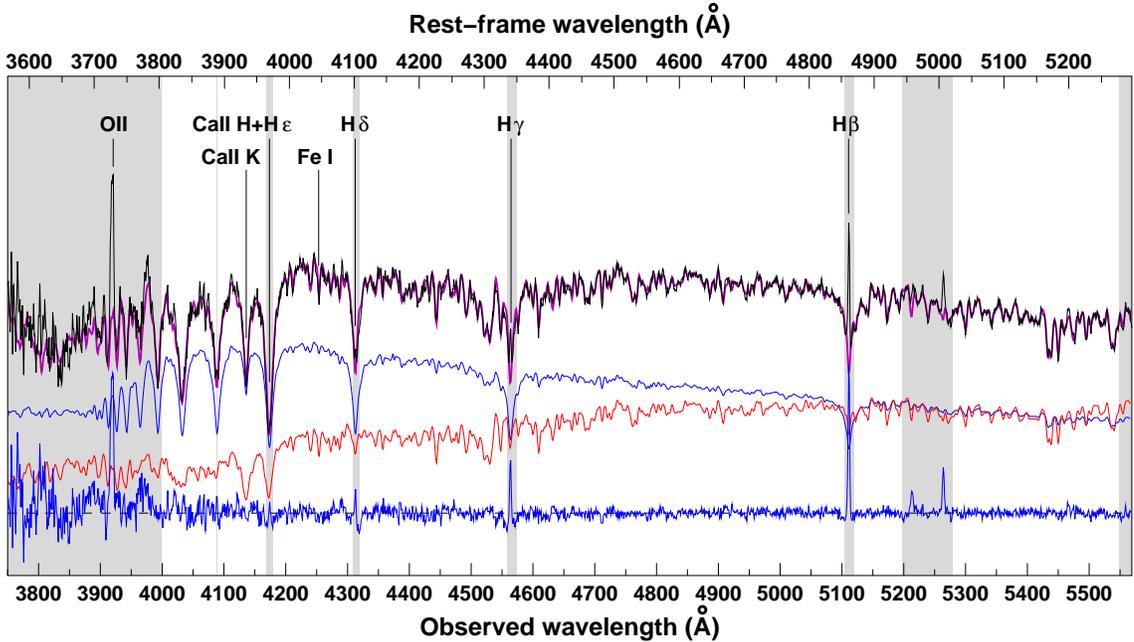}
\caption{The integrated spectrum (black curve) and full population
synthesis fit (magenta) for the NE side of the disc. The curves are
coloured as in Fig~\ref{ssplower}. The line indices discussed in the
text are indicated.}
\label{ssptop}
\end{figure*}

\subsubsection{Recent star formation history}

Having seen that the distribution of dust obscuration and ongoing
star formation is strongly asymmetric, being significantly reduced in
the NE side of the disc, while the SW side shows an ongoing highly
obscured starburst, we now examine the three regions of the galaxy
indicated in Fig.~\ref{fsf24} (SW disk starburst, central bulge starburst,
NE disk) separately.
To determine the ages of stellar populations, we fit the spectrum with
a non-negative linear combination of 40 SSPs, this time with
attenuation of the stellar component by dust also left as a single
free parameter (we do not consider selective attenuation for the
youngest stellar populations), resulting in the best-fit model for the
stellar spectrum. This is then subtracted, leaving the residual noise
and emission spectrum. To achieve the signal-to-noise ratio needed for
this analysis, we added the spectra within each of the three regions
indicated in Fig.\ref{fsf24}.

In the {\it SW disc starburst region} (Fig.~\ref{ssplower}), the
shape of the continuum, together with the weak 4000{\AA} break
($d_{n}4000{=}1.09$), indicate that the spectrum is dominated by young
stars. This is confirmed if we split our best-fit model into its
contribution from young (${\le}1$\,Gyr; thin blue curve) and old
(${>}1$\,Gyr; thin red curve) stars. Young stars contribute $\sim$60\%
of the blue luminosity and ${\sim}3$.5\% of the stellar mass in this
region. Of this young (${\le}1$\,Gyr) component, 35\% of the stellar
mass and 60\% of the luminosity comes from stars younger than
${\sim}1$00\,Myr. This implies a starburst characterised by a
$\sim5\times$ increase in the SFR over the last $\sim$100\,Myr. This
burst is ongoing as implied by the very strong (despite being highly
obscured with A$_{V}{\sim}1$.9--2.6; Fig. 10) H$\alpha$ emission in
this region with EW(H$\alpha){\sim}120${\AA}.
Further Balmer emission lines are visible in Fig.~\ref{ssplower}, from
H$\beta$ all the way up to H$\zeta$.  The old stellar component (red
curve) can be simply characterized as a single 10-15 Gyr old
population.

In the {\it central bulge region} (see
Fig.~\ref{fspect})\footnote{Since the results of the present fit are
qualitatively similar to those shown in Fig.~\ref{fspect}, we refer to
that figure in this analysis.}, the shape of the continuum appears
dominated by an old stellar population. The splitting of our model
stellar populations into old and young, shows that young stars
contribute just ${\sim}3$0\% of the luminosity and 1.0\% of the
stellar mass. Moreover, young stars are predominately of age
${\lesssim}100$\,Myr, as also suggested by the overall smooth appearance of
the continuum from this component.

In the {\it NE disc region} (Fig.~\ref{ssptop}), the stellar
continuum is dominated by young (${\le}1$\,Gyr) stars, contributing to
$\sim$60\% of the luminosity and ${\sim}4$\% of the total stellar
mass. Our stellar population modeling of the young component suggests
multiple populations spread over the full range 60--1000\,Myr with
just ${\sim}5$\% of the stellar mass in this young component coming
from stars younger than 100\,Myr. This suggests that the average SFR
over the last ${\sim}10$0\,Myr is ${\sim}2{\times}$ lower than that
averaged over the last Gyr.

To better understand this situation, we compare the stregths of [O{\sc
ii}] and H$\beta$ emission lines, which trace the ongoing star
formation, to the H$\delta$ absorption line, tracing the
intermediate-age ($\sim$1\,Gyr) A-type stars.  We measure EW([O{\sc
ii}])=-5.2$\pm$0.06{\AA},\\ EW(H$\delta){=}5.24{\pm}0.$08{\AA} and
EW(H$\beta$)=-2.19$\pm$0.06{\AA}. \\ This combination of line strenghts
places the NE disc on the boundary between the weak post-starburst
(k+a) and e(a) (post-starbust with residual star-formation) classes\\
\citep[see][]{dressler99}.  The mean relation between EW(H$\delta$)
and EW([O{\sc ii}]) derived by \citet{dressler04} for normal
star-forming galaxies predicts an
EW(H$\delta$)$\sim$2.0\,{\AA}\footnote{with 95\% of normal
star-forming galaxies lying in the range 0.0--4.0\,{\AA}} for a galaxy
with EW([O{\sc ii}])=-5\,{\AA}. The fact that our observed level of
H$\delta$ absorption is somewhat higher than expected again indicates
that the current SFR is significantly lower than that averaged over
the previous Gyr.

As a further probe of the ongoing-to-recent SFRs we derived the Rose
Ca{\sc ii} index \citep{R85}, which is defined as the ratio of the
counts in the bottoms of the Ca{\sc ii} H+H$\epsilon$ and Ca{\sc ii} K
lines. \citet{R85} showed that the Ca{\sc ii} index is constant at
${\sim}1.2$ for stars later than about F2, but decreases to a minimum
of ${\sim}0.4$ for A0 stars as the Ca{\sc ii} lines weaken, before
increasing again as the H${\epsilon}$ line fades. Furthermore,
\citet{leonardi} combined the Ca{\sc ii} index to the
H$\delta$/Fe\,{\sc i}\,$\lambda$4045 index to determine both the
duration and the strength of starbursts.  We measure a Ca{\sc ii}
index of 0.635 and a H$\delta$/Fe\,{\sc i}\,$\lambda$4045 index of
0.745.  These values confirm that our spectrum is dominated by A-type
stars and, more importantly, show that we are observing a starburst
which lasted for $\sim$0.3\,Gyr and which has been just shut down
\citep[cf. Fig. 4 of][]{leonardi}.

In {\it summary}, the total SFR from the H$\alpha$ flux amounts to
7.2$\pm$2.2\,M$_{\odot}$\,yr$^{-1}$ fully consistent with the previous
estimates from UV and FIR measurements. The star formation across the
galaxy mainly takes place in two regions - in the central bulge region
and $\sim$ 12\,kpc SW from the centre. In the central bulge region
there is an intense ongoing, heavily obscured star formation
accounting for $\sim$30\% of the total SFR of SOS\,114372. In the SW
starbursting region, which accounts for $\sim$20\% of the total SFR,
our data imply a starburst characterised by a $5{\times}$ increase in
the SFR over the last ${\sim}10$0\,Myr. This burst is ongoing. In the
NE disc, our data still show ongoing star formation, but significantly
lower than in the rest of the disc. The full spectral modeling and the
line strengths show that we are observing a 0.3 Gyr starburst
immediately after it has been shut down.

\section{Origin of the extraplanar gas}
\label{dis}

The most prominent and notable physical characteristic of
SOS\,114372 is the extraplanar ionized gas extending out to a
projected distance of 13\,kpc NW from the disc over its full extent
following the rotation of the disc. This gas is resolved into a
complex of compact H$\alpha$-emitting knots and faint filamentary
structures and shows a high fraction (0.4-0.8) of shock
excitation. All this supports a scenario in which gas is being driven
out from the galaxy disc. The high levels of dust attenuation with
$A_V{\sim}1$.9\,mag also found associated with the extraplanar gas,
suggests that the dust is stripped and dragged away from the disc
together with gas, in agreement with the reduced dust disc observed in
HI-deficient Virgo galaxies by \citet{CDP10}.

Causes of large-scale outflows of gas from a galactic disc may be
galactic winds produced by AGN or starbursts, tidal interaction or
ram-pressure stripping. In the following we will outline the arguments
that lead to the idea that RPS is actually the origin of the
observed gas outflow.

\subsection{Galactic winds}
\label{GW}

We notice that the outflow takes place almost uniformly all over the
disc and is observed only in one direction (NW, see
Figs.~\ref{foutflow} and~\ref{halpha}). Since it extends in projection
several kiloparsecs outside of the disc, dust attenuation could not
hide a possible (symmetric) component on the SE side. The observed
outflow is therefore intrinsically asymmetric. Galactic winds,
instead, produce bipolar outflows, with a structure ranging from
biconical to egg-shaped, originating from a nuclear starburst or AGN,
and extending on both sides of the disc \citep[][and references
therein]{VCB05}. We
therefore exclude a galactic wind as the cause of the bulk of the gas
outflow observed here.

Nevertheless, we don't exclude that a galactic wind could actually be
in place close to the nucleus, where star formation is taking place at
a rate of $\sim$2\,M$_{\odot}$yr$^{-1}$ and the H$\alpha$ image shows
hints of gas escaping from the centre of the SE disc (see
Fig.~\ref{halpha}). Let us consider the area with very high values of
the velocity dispersion (${\gtrsim}$ 120 km s$^{-1}$, Fig.~\ref{fGVF})
located along the minor axis SE from the galaxy centre. We notice that
this area has a roughly triangular shape, with a vertex near the
galaxy centre. The minor axis radial velocity profile
(Fig.~\ref{fGRC}) shows that the gas is moving towards the observer,
with a speed steadily rising up to $\sim$100\,km\,s$^{-1}$ at $\sim
3$\,kpc from the centre.  The high values of $\sigma$ suggest the
presence of different components of gas with different radial
velocities, so that the measured $\sim$100\,km\,s$^{-1}$ is an average
velocity (weighted by the H$_{\alpha}$ flux).

All these features, together with the fact that the radiation here is
shock-dominated (Sect.~\ref{PPG}), suggests that the triangular region
SE from the centre may be a semi-cone produced by the nuclear
starburst \citep{VCB05}. The absence of the companion semi-cone in the
NW direction could be explained by dust absorption if the disc is
oriented with the SE side furthest from us. We will add further
support for this orientation in Sect.~\ref{simul}. Given the
inclination of the galaxy and the apparent extension of the wind cone
of 6.5\,arcsec from the centre, and allowing for large uncertainties
due to poor spatial resolution, we infer that the cone extends
$\sim$ 6\,kpc above the disc, within the typical sizes of observed
galactic winds \citep{VCB05}.

As already noted, the area SE from the nucleus is the most
dust-obscured (excluding the nuclear region itself;
Fig.~\ref{fatt}). This would be easily explained by a large amount of
dust collected in the starbursting region and entrained in the wind,
which is a rather common circumstance in galactic winds \citep[][and
references therein]{HHS05}.

\subsection{Tidal interaction}
\label{TS}

The best candidates for a major tidal interaction are the cluster
members SOS\,114493 and SOS\,112392 (Fig.~\ref{SOS} and
~\ref{duegalassie}) with redshifts $z= 0.050$ and $z= 0.046$ and
$K$-band magnitudes 13.49\,mag and 12.61\,mag, respectively. These two
galaxies are at projected distance from SOS\,114372 of $\sim$45\,kpc
and $\sim$55\,kpc respectively, and their stellar masses are
$\mathcal{M}_{114493}$=$2{\cdot}10^{10}$M$_\odot$ and
$\mathcal{M}_{112392}$=$7{\cdot}10^{10}$M$_\odot$ (Paper\,I). We
observe no tidal bridge connecting SOS\,114372 to either of these
objects nor any signs of disturbance in their $R$-band images, as is
shown in Fig.~\ref{duegalassie}. The direction of the gas outflow from
SOS\,114372 does not appear to be related to any of these galaxies.
Furthermore, no obvious sign of perturbation is present in any of SSC
galaxies within 200\,kpc projected radius from SOS\,114372.

\begin{figure}
\includegraphics[width=84mm]{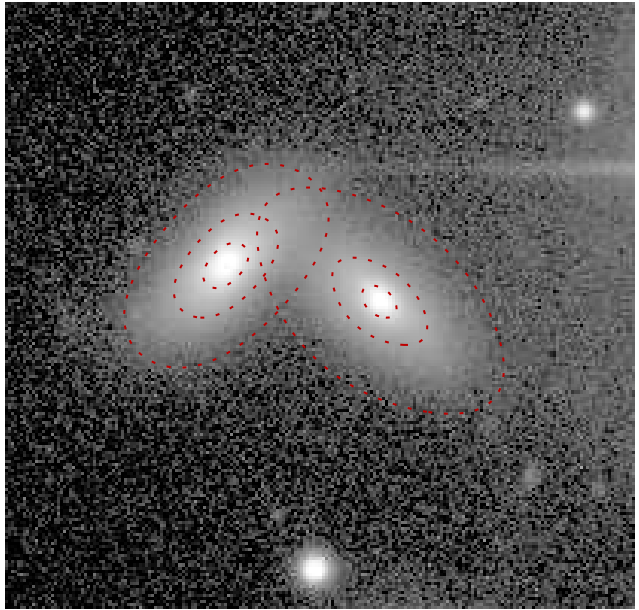}
\caption{Close-up of SOS\,114493 (left) and SOS\,112392 (right) from a
  ESO-2.2m WFI deep R-band image. The intensities are stretched by a
  logarithmic scaling to emphasize the lowest surface brightness regions. The
  deep-red ellipses are drawn from the models derived from a
  simultaneous fit to the two galaxies using two S\'ersic laws. The
  fit was performed with GALFIT \citep{PHI10}. The surface brightness
  of both galaxies smoothly follow the models without any appreciable
  sign of perturbation. One side of this figure corresponds to
  $\sim$45\,kpc.}
\label{duegalassie}
\end{figure}

As already noted in Sect.~\ref{GSK}, the stellar velocity field of
SOS\,114372 displays no sign of disturbances, indicating that the
galaxy is not suffering any significant external gravitational perturbation.
The lack of a tidal tail, the undisturbed stellar disc, the existence
itself of the gas disc, and the absence of perturbed neighbours make
us exclude the occurrence of tidal interaction, unless it is at a very
early stage, with the closer galaxy (at a distance anyway $>$ 45\,kpc)
just starting its first approach.
Numerical simulations \citep{KKS06,KKS07}, however, do not
predict gas outflow at this stage, either for two galaxies of the same
mass (as in the case of SOS\,114372--SOS\,112392) or for the
unequal-mass case (as for SOS\,114372--SOS\,114493).

The next candidate for a tidal interaction is the faint object located
SW of SOS\,114372 (SOS\,115228, see Fig.~\ref{SOS} right
panel). Although we do not have redshift information for this galaxy,
it appears much brighter in our narrow-band H$\alpha$ image than in
the continuum, which would suggest that it is a starbursting dwarf
galaxy at a redshift within a few hundred km\,s$^{-1}$ of
SOS\,114372. As tidal torques are effective in channelling rapid gas
inflows capable of fuelling nuclear starbursts, it seems reasonable
that an interaction between SOS\,114372 and SOS\,115228 could be
behind the observed starburst in the dwarf galaxy. It is much less
likely that this interaction could induce the large-scale gas outflow
in SOS\,114372 however. Firstly, the $K$-band magnitudes of the two
galaxies differ by $\sim$ 5\,mag, and we found a factor about 360
difference in stellar masses (Paper\,I). This makes it unlikely that
this object can have such a large-scale effect on the gas of its much
more massive neighbour. Secondly, simulations of unequal mass mergers
predict gas streams connecting the two galaxies (see Fig.~6 of
Kronberger et al. \citeyear{KKS06}), while we observe the gas outflow
in the opposite direction. We find no evidence at all for a tidal
bridge connecting the two galaxies, although \citet{KKF09} find that
in clusters the impact of pressure from the ambient ICM can destroy
these tidal bridges, as well as further enhance the levels of star
formation in the interacting galaxies. Notice, however, that the
simulated galaxies considered by \citet{KKS06} to investigate the
unequal mass merger have at most a mass ratio equal to 8:1. It may be
that the notable starburst on the SW-edge of SOS\,114372 is caused by
such a combination of the tidal interaction and ram pressure
stripping. Finally, as already noted in Sect.~\ref{Ha_morph}, it is
possible that SOS\,115228 is at a completely different redshift from
SOS\,114372.

All the above considerations and most importantly the symmetric
stellar distribution of the three brighter galaxies lead us to
conclude that tidal interaction is not the cause of the observed gas
outflow.

\subsection{Ram-pressure stripping} 
\label{SRPS}

The large-scale gas outflow detected by the IFS data is resolved in
the narrow-band H$\alpha$ imaging into a complex of compact knots of
H$\alpha$ emission, often attached to faint filamentary strands which
extend back to the galactic disc (Fig.~\ref{halpha}). Similar
complexes of H$\alpha$ or UV-emitting knots and filaments have been
seen for numerous other galaxies
\citep[e.g.][]{GBM01,CMR07,YYK10,VSC10,FGB12}. The most notable
aspects of these systems are that all the affected galaxies are
located in cluster environments, and that the outflow is one-sided,
often characterized as a cometary tail of ionized gas directed away
from the cluster centre. These two aspects both point to a
cluster-specific process in which gas is stripped from an infalling
galaxy by its passage through the ICM.

The H$\alpha$ velocity field of SOS\,114372 can be compared with
those of nearby galaxies affected by RPS such as NGC\,4438
\citep{CCB05,VSC09} and NGC\,4522 \citep{VMA00} in the Virgo cluster.
Both galaxies show one-sided extraplanar gas following the rotation of
the disc, as for SOS\,114372. The velocity field of NGC\,4438 is
characterized by filamentary structures extending out of the disc. If
present, such filaments could not be resolved at the higher redshift
of SOS\,114372 with our IFS data. Remarkable analogies are found in
the H$\alpha$ images of the two galaxies (compare Fig.~2 of Kenney \&
Koopmann \citeyear{KK99} with our Fig.~\ref{halpha}). \citet{KK99}
observed that the H$\alpha$ emission arising from both \HII\, regions
and diffuse emission is organized in filaments extending more than
3\,kpc from the outer edge. The difference between the two galaxies is
in the truncation radius - NGC\,4522 has a truncated gas disc at
radius of 3\,kpc, while in the case of SOS\,114372 the ionized gas
extends up to 13\,kpc. This difference can be related to different
'ages' of RPS, different orbits, ICM densities and ICM wind angles. We
also notice that there are intrinsic differences between the two Virgo
galaxies and that studied in this article. SOS\,114372 has an absolute
$B$-band magnitude of M$_B=-21.48$, significantly brighter than
M$_B=-20.04$ and M$_B=-18.10$ of NGC\,4438 and NGC\,4522, respectively
(assuming 16\,Mpc as distance of the Virgo cluster and the magnitudes
from Chung et al. \citeyear{CvG09}). This shows that we are comparing
galaxies of different masses. Furthermore, NGC\,4438 is located close
to the Virgo cluster centre (280\,kpc from M87, $r\sim$0.3$r_{200}$)
while NGC\,4522 is at $r\sim r_{200}$ \citep{UWS11} and SOS\,114372 is
at $r\sim$1\,Mpc ($\sim$0.5$r_{200}$) from the centre of the rich
cluster A\,3558. It is therefore likely that these galaxies are
experiencing different ICM-ISM interactions. The common point between
the velocity fields of NGC\,4522, NGC\,4438, and SOS\,114372 is that
the extraplanar regions are still dominated by rotation.

Hydrodynamical simulations show that ram-pressure and viscous
stripping of spiral galaxies produce narrow wakes of H{\sc i} gas that
can extend up to 100--200\,kpc and last 500\,Myr \citep{TB10}. These
simulations predict a wide range of densities and temperatures within
these tails, leading to the in-situ formation of pressure-supported
clouds dense enough to form stars. This star-formation plus localized
heating via compression and turbulence can light up the H{\sc i} tail
with H$\alpha$ emission, with intensities and morphologies
\citep{TB12} similar to those observed in SOS\,114372. \citet{TB12}
note that this star formation in the tail does not occur due to
molecular clouds that were formed within the galaxy itself and
stripped wholesale, but rather because the relatively low-density gas
that has been stripped can cool and condense into dense clouds in the
turbulent wake. They also indicate that higher ICM pressures create
more dense clouds in the wakes, resulting in higher SFRs.

\begin{figure}
\includegraphics[width=84mm]{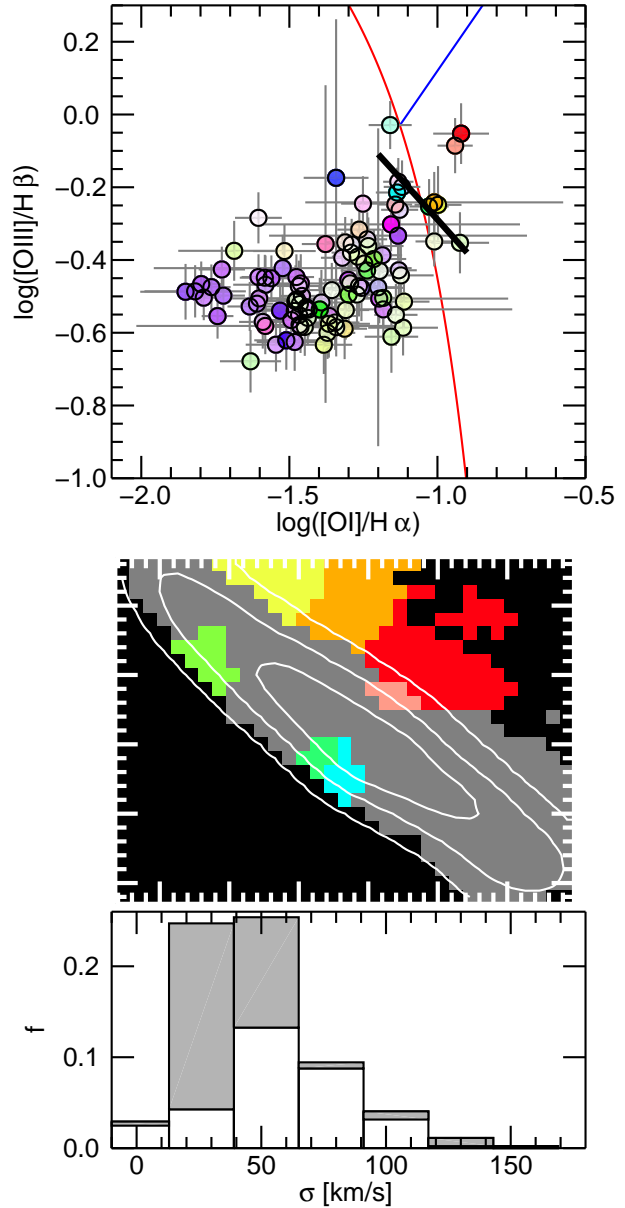}
\caption{{\it Upper panel}: The flux ratios in the different galaxy
regions as in Fig.~\ref{fbpt3}. The black curve corresponds to the
shock and photoionization model where 50\% of the flux is due to
shock-ionized gas. {\it Central panel}: The galaxy regions dominated
by the contribution of shock-ionized gas colour-coded as in
Fig.~\ref{fvsk}. {\it Lower panel}: Normalized distribution of the
velocity dispersion values. The white coloured histogram refers to the
regions dominated by the contribution of shock-ionized gas, while the
gray histogram refers to the ``low'' shock-ionized gas region. The
regions characterized by shock-like line ratios tend to have higher
velocity dispersions.}
\label{fvsk}
\end{figure}

Nevertheless, star formation cannot be the only origin of the
H$\alpha$ emission coming from the outflowing gas. In Fig.~\ref{fvsk}
(upper panel) we focus again on the right panel of Fig~\ref{fbpt3}, which
plots the [O{\sc iii}]/H$\beta$ versus [O{\sc i}]/H$\alpha$
diagnostic. There are seven galaxy regions where shock excitation
contributes more than 50\% of the emission. Their flux ratios are
located to the right side of the black curve in upper panel of
Fig.~\ref{fvsk} and are coloured in the central panel. These regions
correspond to the gas outflow and to three regions in the SE edge of
the galaxy. In all these regions, except the easternmost one (shown in
light green), we also measure strong dust extinction. The normalized
distribution of the velocity dispersion measured for these regions
(white coloured histogram in Fig.~\ref{fvsk}) is shifted to higher
values with respect to the overall distribution. We expect the
velocity dispersion produced by the shocks to be of the same order as
the shock velocity. If $n_{ICM}$ and $n_{ISM}$ are the number
densities of the ICM and ISM respectively and if $v_o$ is the relative
velocity of the two gas components, then the shock velocity is
$v_s$=$\sqrt(n_{ICM} / n_{ISM}) \times v_o$. If $n_{ICM} / n_{ISM}
\sim$ 10$^{-2}$--10$^{-3}$ and $v_o \sim$ 1000 \,km\, s$^{-1}$ then
$v_s \sim$30-100\,km\, s$^{-1}$, which is what we observe. The high
velocity dispersions of some non-shocked regions may be well explained
by the complicated radial motions produced by ram pressure and by the
fact that our resolution element intercepts gas clouds for depths of
$\gtrsim 1/cos\ i \sim$ 7 kpc.

From the distribution of the regions with higher velocity dispersion
and shock-ionized gas across the galaxy, we can infer that SOS\,114372
is affected by ram pressure which compresses and strips the gas out of
the galaxy, forming a tail of turbulent shock-ionized gas and
dust. Part of this gas can cool and condense into dense clouds
observed in H$\alpha$.

The influence of RPS on the internal kinematics has been investigated
through N-body/hydrodynamical simulations by \citet{KKU08b}
demonstrating that the gas velocity field shows clear evidence of
disturbances after a few tens of Myr from the onset of RPS. In
particular, the rotation curve can be characterized by asymmetry in
the external gas disc while the inner disc remains undisturbed. A
measure of the disturbance is given by the asymmetry parameter of the
rotation curve as defined by \citet{DGH01}. We estimate an asymmetry
parameter of about 20\% for the SOS\,114372 major axis radial velocity
profile, consistent with what estimated by \citet{KKU08b} for
RPS. Moreover the major axis velocity profile of SOS\,144372
(extending to ${\sim}3.8{\times} r_d$; Fig.~\ref{fGRC}) is very
similar to the gas rotation curve derived for the model galaxies after
50\,Myr of ram pressure acting face-on (Fig.~2 of Kronberger et
al. \citeyear{KKU08b}) extending to $4{\times}r_d$.

SOS\,114372 is amongst the brightest SSC galaxies in the mid-infrared
with one of the highest SFRs measured for the whole sample (see
Paper\,II). The galaxy centre and the SW region of the disc contribute
$\sim$30\% and $\sim$20\% respectively to the total SFR measured by
the H$\alpha$ luminosity (i.e. current SFR). The location of the
star-forming regions in the galaxy agrees with the predictions of the
simulations of RPS where the newly formed stars are preferentially
located in the compressed central region of the galaxy and/or in the
stripped gas behind the galaxy \citep{KKF08a} depending on the
surrounding gas density \citep{KSS09}. Sanderson \& Ponman
(\citeyear{SP10}) estimated an ICM density of $\rho_{ICM}=1.3 \times
10^{-28}$g\,cm$^{-3}$ at the projected distance of the galaxy from the
cluster core. This density implies that all the newly formed stars are
confined into the disc either because of the low ICM density or
because the RPS has started less than 100\,Myr ago \citep{KSS09}. The
starburst detected in the SW disc region produces a ${\sim}5{\times}$
increase in the SFR over the last ${\sim}10^{8}$\,yr and could be
ram-pressure induced. Also in the central bulge region, whose spectrum
is clearly dominated by old (${>}1$\,Gyr) stars, we measure a
non-negligible contribution from young (${\lesssim}10^{8}$\,yr) stars, but
this star-forming region is most probably related to the presence of
the bar (see Sect.~\ref{con}).

The asymmetry observed in the SFR and stellar population properties
across the galaxy disc can be explained in a RPS scenario where the
ram pressure contributes to the star formation quenching via the
impoverishment of the gas reservoir, while inducing the ongoing
starburst in the highly obscured SW side. Different stripping
efficiencies depending on the surface density of the stripped gas
clumps are also found by Vollmer et al. (\citeyear{VSC09}).

We can thus conclude that all the observed properties of SOS\,114372
can be accommodated in the RPS scenario.

\subsection{Environmental conditions for\\ ram-pressure stripping}
\label{feas}

Now we consider the ICM properties in order to check if RPS can be
effective for this galaxy.

For A\,3558, Sanderson \& Ponman (\citeyear{SP10}) derived the dark
matter density, gas density ($\rho_{ICM}$) and the gas temperature
profiles by fitting Ascasibar \& Diego (\citeyear{AD08}) models to the
{\it Chandra} X-ray data. In these models, the dark matter is
described by a Hernquist (\citeyear{H90}) density profile, which
provides simple analytic expressions for the mass profile,
gravitational potential and escape velocity. From this, we can derive
the ram pressure, $P_{ram}{=}\rho_{ICM}v_{orb}^2$, acting on galaxies
as function of cluster-centric radius and orbital velocity (Haines et
al. \citeyear{HBM11c}). For the cluster-centric radius we assume the
projected distance of 0.995\,Mpc. As a lower limit for the orbital
velocity, we assume the galaxy's line-of-sight velocity relative to
the cluster's systemic velocity $V_{los}\sim$830\,km\,s$^{-1}$, For
the upper limit we assume the escape velocity at the cluster-centric
distance which, from the Henrquist model for A\,3558, is
$\sim$2600\,km\,s$^{-1}$. At 0.995\,Mpc from the cluster centre
$\rho_{ICM}\simeq$1.3$\cdot$10$^{-28}$\,g\,cm$^{-3}$ and the ram
pressure ranges from $\sim$9$\cdot$10$^{-13}$\,dyn\,cm$^{-2}$ to
$\sim$9$\cdot$10$^{-12}$\,dyn\,cm$^{-2}$. This is the regime of weak
to moderate ram pressure as defined by Roediger \& Hensler
(\citeyear{RH05}) who, by means of high resolution 2D hydrodynamical
simulations, demonstrated that ram pressure effects can be observed
over a wide range of ICM conditions. In particular, in high density
environments RPS severely truncates the gas disc of $L^\star$
galaxies, while in low density environments, where moderate ram
pressure is foreseen, their gas disc is clearly disturbed and bent
\citep{RH05} as we observe. The gas discs of these galaxies can be
truncated to 15-20\,kpc in the first 20--200\,Myr of RPS. To check if
the expected ram pressure is able to strip the gas from the disc of
SOS\,114372, we estimate the gravitational restoring force per unit
area $(d\phi/dz)\sigma_{gas}=2\pi G\sigma_{disc}\sigma_{gas}$, with
$\sigma_{star}$ and $\sigma_{disc}$ the star and gas surface density,
respectively. From the $K$-band photometry, we derive the inclination-
and dust attenuation-corrected surface brightness at 20\,kpc (where
the gas disc is bent) of 4.4\,L$_\odot$pc$^{-2}$. From the $K$-band
luminosity and the stellar mass (see Table~\ref{SOS114372}), we obtain
M/L$_K\sim$2, therefore the stellar mass surface density is
8.8\,M$_\odot$pc$^{-2}$. Assuming a typical $\sigma_{gas}\sim
10$\,M$_\odot$pc$^{-2}$, the restoring force per unit area is
$1.7\cdot 10^{-12}$\,dyn\,cm$^{-2}$ which compared with the estimated
ram pressure (9$\cdot$10$^{-13}$--9$\cdot$10$^{-12}$\,dyn\,cm$^{-2}$)
shows that it can be effective in removing the gas from the disk if
$v_{orb}\ge 1100$\,km\,s$^{-1}$. We assume that the galaxy is coming
in radially, such that the projected cluster-centric distance is the
same as the actual 3D one. If the galaxy is at a larger 3D distance,
then the local density will be lower, as will the wind velocity, and
thus the ram pressure could be overestimated although a moving ICM
and/or local overdensities can enhance it.

Crowl \& Kenney (\citeyear{CK06}, \citeyear{CK08})
conducted a survey of ten Virgo galaxies with the SparsePak IFS. They
found clear evidence of RPS-induced star formation within the
truncation radius and a passive population beyond it. Their estimates
of the stripping timescales showed that Virgo galaxies can be stripped
both within and beyond the cluster core. This is direct evidence that
the ICM is neither static nor homogeneous. That a galaxy's ISM can be
affected by the environment also in intermediate- to low-density
regions (0.6--1.0\,Mpc from the cluster cD galaxy) in the Virgo
cluster was also concluded by \citet{CGK07} observing the HI tails of
seven spiral galaxies.

A non homogeneous ICM is expected for non virialized merging and
post-merging clusters. Dynamical analyses \citep{BPR98} and X-ray
observations (Bardelli et al. \citeyear{BZM96}; Akimoto et
al. \citeyear{AKF03}) of A\,3558 indicate this cluster is the result
of a cluster-cluster merging and/or a group infall. More recently,
combining {\em XMM-Newton} and $Chandra$ data, \citet{RGM07} were able
to detect a cold front in the A\,3558 ICM probably caused by the
sloshing of the core induced by the perturbation of the gravitational
potential associated with a past merger. This cold front is expanding,
perturbing the density and the temperature of the ICM. From their
Fig.~3 it is clear that the cold front is moving in the direction of
SOS\,114372 and that it is $\sim$5\,arcmin ($\sim 300$\,kpc in
projection) from the galaxy. Unfortunately, the X-ray observations do
not include the region of SOS\,114372, which is planned to be observed
in future surveys (see Hudaveri, Bozkurt \& Ercan \citeyear{HBE10}).

Taking into account all the above arguments, we conclude that the
environment of our galaxy is favourable to RPS despite its large
distance from the cluster centre.

\section{N-body/hydrodynamical simulations of ram-pressure stripping}
\label{simul}

To confirm the scenario of RPS and to constrain its physics, we
performed N-body/hydrodynamical simulations. Our aim was to reproduce
as closely as possible the velocity field of the gas. We remark that
in the real system there are factors that cannot be accounted for in a
simulation, such as the spatial distribution of the dust, which
may affect the observed kinematics and distribution of the gas. The
interaction of the ICM with the ISM may also be complicated by their
clumpiness, not to mention the action of the bar. We therefore don't
expect the simulations to provide a description of the details of the
gas kinematics, but aim to obtain an overall agreement with the
observations helping us to constrain the ICM wind angle and velocity
and the `age' of RPS.

\begin{table*}
  \centering
\caption{{\bf The model galaxy.} Properties of the initial conditions
  of the model galaxy used. The particle numbers, mass resolution and
  cumulative mass are shown for disc scale length corresponding to
  10\% of the disc radius both for low- and high-resolution
  simulations (LR, HR).}
  \begin{tabular}{rccc}
&&&\\
    \hline \hline & \# of particles & mass resolution & total mass\\ &
     & $\left[ M_\odot/particle \right]$ & $\left[ M_\odot \right]$ \\
     \hline 
HR&&&\\
DM halo \vline & $3\times10^{5}$ & $3.7\times10^{6}$ &
     $1.1\times10^{12}$ \\ 
gas disc \vline & $2 \times10^{5}$ &
     $2.9\times10^{4}$ & $5.9\times10^{9}$ \\ 
stellar disc \vline &
     $2\times10^{5}$ & $2.6\times10^{5}$ & $5.3\times10^{10}$ \\
     
stellar bulge \vline & $1\times10^{5}$ & $1.7\times10^{5}$ &
     $1.7\times10^{10}$ \\ \hline \\
LR&&&\\
DM halo \vline & $3\times10^{4}$ & $3.7\times10^{7}$ &
     $1.1\times10^{12}$ \\ 
gas disc \vline & $2 \times10^{4}$ &
     $2.9\times10^{5}$ & $5.9\times10^{9}$ \\ 
stellar disc \vline &
     $2\times10^{4}$ & $2.6\times10^{6}$ & $5.3\times10^{10}$ \\
     
stellar bulge \vline & $1\times10^{4}$ & $1.7\times10^{6}$ &
     $1.7\times10^{10}$ \\ \hline \\
  \end{tabular}

  \label{tab:ic_galaxy}
\end{table*}

\subsection{Simulations}
An important aspect of the use of simulations is that, at least in the
early phases of RPS, the direction of the ICM wind cannot be inferred
from the apparent direction of the gas outflow, because the motion of
the gas is a complex combination of the effects of the ram pressure
and the gravitational field of the host galaxy (see also Roedigger \&
Br\"uggen \citeyear{RB06}). Clarke et al. (\citeyear{CKP98})
introduced a geometrical scheme (see their Fig. 1) to study the jets
in Seyfert galaxies. In that scheme, the angle between the jet and the
normal to the disc is constrained by the apparent jet outflow angle
and the inclination of the galaxy. In a different framework, that
scheme was adopted by Abramson et al. (\citeyear{AKC11}) to study the
geometry of RPS, substituting the jet direction with the ICM wind
direction.

Unfortunately, we cannot adopt Clarke et al.'s scheme for two
reasons. The first reason is that the rotation of the galaxy breaks
one important symmetry that lies at the ground of the applicability of
that scheme. The second reason is that not only the direction of the
outflow does not indicate the direction of the wind, but also the
direction of the wind with respect to the disc cannot be assumed {\it
a priori} (see Appendix\,A). We therefore considered all the possible
values for the angle $\beta$ between the galaxy rotation axis and the
ICM wind. We ran simulations from RPS close to edge-on
($\beta=85^{\circ}$) to close to face-on ($\beta=15^{\circ}$) at
intervals of 10$^{\circ}$ (there was no need to explore values closer
to face-on or to edge-on). For what concerns the ICM wind velocity, we
set as the lower limit the galaxy's line-of-sight velocity relative to
the cluster's systemic velocity, $V_{los}\sim$830\,km\,s$^{-1}$, and
as the upper limit the escape velocity, $\sim$2600\,km\,s$^{-1}$
(derived from the Hernquist model for A\,3558).

We explored the following values for the ICM wind velocity
($V_{wind}$=830, 1100, 1400, 1700, 2000, 2200, 2400,
2600\,km\,s$^{-1}$). We also explored two different disc scale heights
corresponding to 10\% and 20\% of the radius $r_d$. In summary, our
grid of models includes 8 values for $\beta$, 8 values for $V_{wind}$
and 2 values for the disc scale height. Our approach was to run
low-resolution simulations for the whole range of parameters and then
select the most promising cases for the high-resolution simulations.

The simulations were done with the cosmological N-body/hydrodynamic
code GADGET-2 developed by Volker Springel \citep{springel05}. The
hydrodynamic part follows an SPH scheme \citep{gingold77, lucy77}
while the gravitational interaction is calculated using a tree code
for the short range force as first introduced by \citet{bh86} and
treePM code based on Fourier techniques for the long range
forces. Furthermore, our version of GADGET-2 includes radiative
cooling \citep{katz96} and recipes for star formation, stellar
feedback and galactic winds according to \citet{springel03}. 

To simulate galaxies undergoing RPS, we use the same simulation setup
as \citet{KSS09}, \citet{KKF08a} and \citet{steinhauser12}. First, we
generate a model galaxy matching the observed properties of
SOS\,114372 (see Sect.~\ref{gal}). The initial conditions are
calculated according to \citet{springel05a} based on the work of
\citet{mo98}. Our model galaxy includes a dark matter halo, a stellar
disc, a gaseous disc and a stellar bulge. The disc scale length of
the galaxy is directly related to the angular momentum of the disc
with the spin as a free parameter. The stellar and gaseous disc have
exponential surface density profiles. For the stellar bulge and the
dark matter halo a \citet{H90} profile is adopted. Important
parameters and properties adopted for our model galaxy are shown in
Table~\ref{tab:ic_galaxy} for both the low- and high-resolution
simulations. 
The total mass is $\mathcal{M}_{tot}{=}1.2{\cdot}10^{12}$M$_\odot$ and
the total (dynamical) and baryonic mass within 20\,kpc are
$\mathcal{M}_{dyn}{=}2.1{\cdot}10^{11}$M$_\odot$ (see Sect.~\ref{GSK})
and $\mathcal{M}_{\star}{=}7.0{\cdot}10^{10}$M$_\odot$,
respectively.
Since the initial gas mass of SOS\,114372 is not known, we assume that
the gaseous phase contains 10\% of the mass of the disc. The initial
disc scale height was set to either 10\% or 20\% of the disc scale
length. 
The disc-to-bulge mass ratio of 0.32 adopted for the model galaxy was
derived from the measured B/D flux ratio of 0.44 taking into account
that $\sim$30\% of the light from the bulge is due to very young stars
contributing only 1\% to its mass (Sect.~\ref{csf}).

Since the model galaxy needs some time to settle to avoid numerical
artefacts, it is evolved in isolation for 1\,Gyr before the ICM wind
is applied. Furthermore, the parameters for generating the model
galaxy were chosen carefully to obtain a stable galaxy. As also found
by \citet{moster11}, the resolution plays an important role. As we
found in the low-resolution runs in which the galaxy consists of
a significantly smaller number of particles than the nominal
$7.5\times 10^5$ value, both the stellar and the gaseous disc were
puffed up, leading even to warps although the galaxy is evolving in
isolation. With the chosen parameters the effect of the resolution was
less dramatic and negligible on the morphology of the stripped gas,
used for the selection of the models which better reproduce the
observations.

To simulate the prevailing environment of \\ SOS\,114372, the model
galaxy is then exposed to an external wind, simulating the ICM. At
variance with many other RPS simulations, the model galaxy itself is
moving and the ICM is stationary. The gas particles representing the
ICM fill a simulation domain with periodic boundary conditions in
order to keep the ICM density stable during the simulation. The mass
resolution of the ICM particles is the same as for the ISM particles
to avoid numerical artefacts. As we use a box size of 850\,kpc, a
large number of particles would be needed leading to a very long
computation time. For this reason, only the inner part of the cube
with size $200\times 200\times 850\,\mathrm{kpc}^3$, where the galaxy
flies through, has the same mass resolution as the ISM. On the
outside, particles with a hundred times higher mass are used. For an
ICM density of $\rho_\mathrm{ICM} = 1.3 \times
10^{-28}\,\mathrm{g}\,\mathrm{cm}^{-3}$ \citep{SP10}, 3.1 million gas
particles are needed. 

The models were run from the onset of the ram pressure to 100\,Myr
after. During this time the ram pressure is kept constant and its
onset is immediate.

\begin{figure*}
\includegraphics[width=150mm]{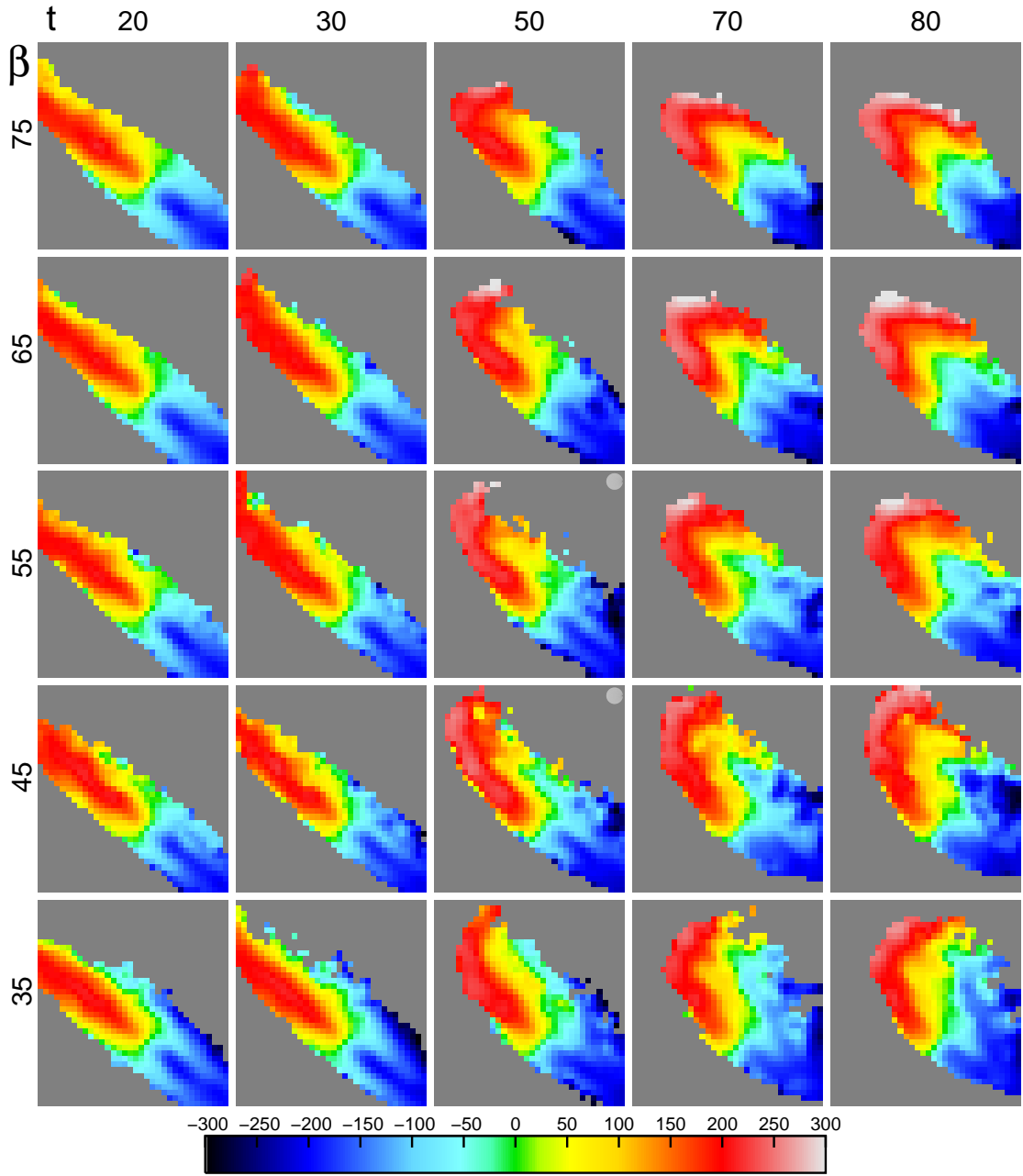}
\caption{Example of a grid of low-resolution models. The wind angle
$\beta$ (in degrees) changes in the vertical direction and the time
since the onset of RPS (in Myr) in the horizontal direction as
indicated. The selected models are marked with a light circle in the
top-right of the thumbnails. Notice that all of the models are shown
for brevity only in one projection(case $B$, see Appendix). $V_{wind}$
is fixed at 1400\,km\,s$^{-1}$. The colour bar gives the values of
radial velocity in km\,s$^{-1}$. Notice the progressive bending of the
disc with time and the progressive truncation of the upper-left disc
with $\beta$.}

\label{simul1}
\end{figure*}

\begin{figure*}
\includegraphics[width=150mm]{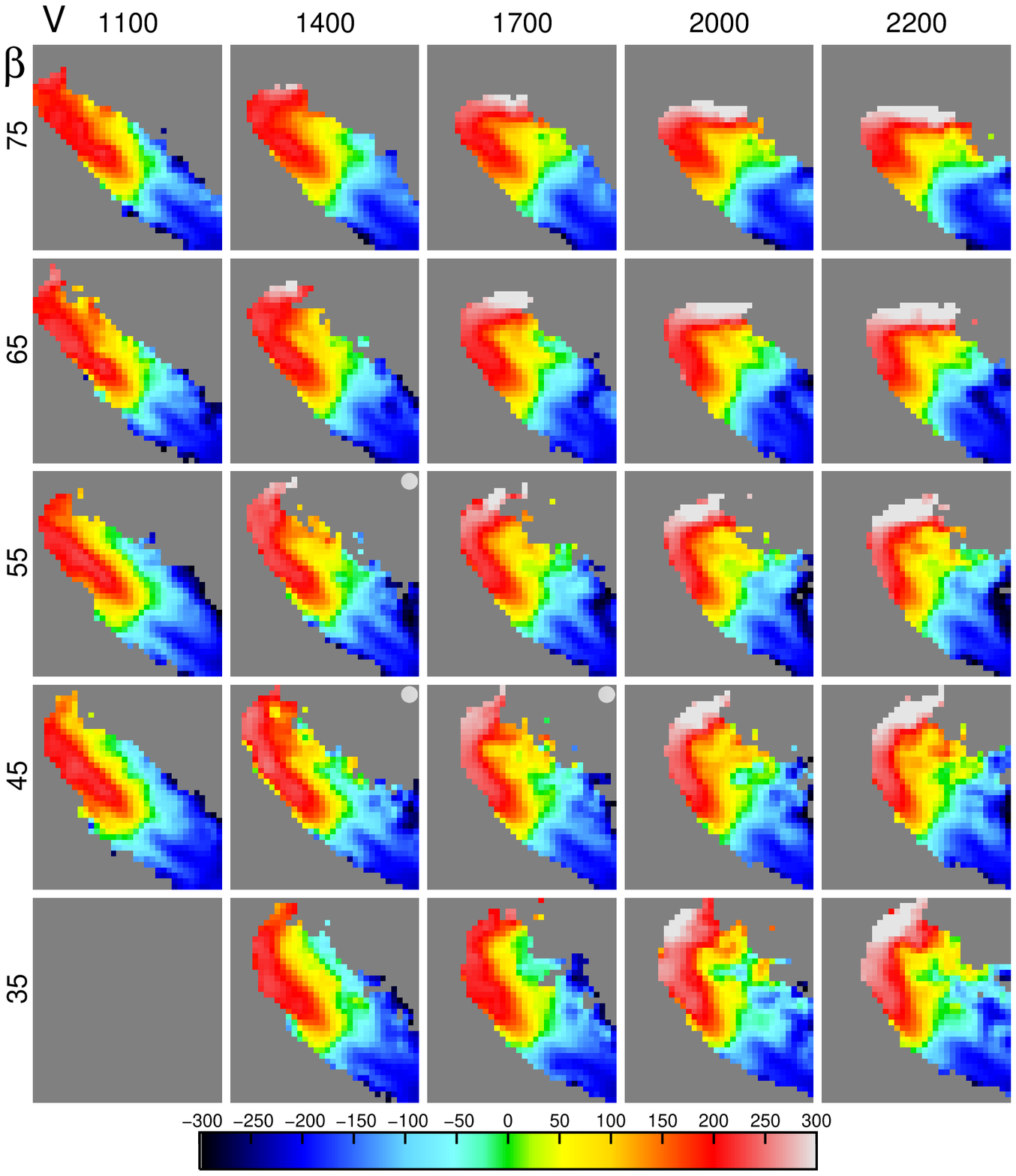}
\caption{Analogous to Fig.~\ref{simul1}, but with the wind velocity
  $V_{wind}$ in place of the time. The time is fixed at 60\,Myr.}

\label{simul2}
\end{figure*}

\subsection{Comparison with the observations}
To compare the models with the data we first accounted for the
constraints on the viewing directions. The first constraint is given
by the inclination angle $i$=82$^{\circ}$ of the galaxy. The other
constraint comes from the requirement that the spatial velocity of the
galaxy $\overrightarrow{V}_{gal}$ projects into the observed
$V_{los}$. This translates into the requirement that the angle $\phi$
between the motion of the galaxy and the line of sight is given by
$cos\ \phi = V_{los}/V_{gal}$. As it is shown in the Appendix, the
consequence of these constraints is that some models can be always
observed from four directions, others from two, and others from two
but only for high values of $V_{gal}$. The issue of the viewing
directions is treated in detail in the Appendix where we also describe
the process leading from the 3-D models provided by the simulations to
the 2-D fields comparable with our data. In brief, to compare the
models with the observations we account for the seeing effect
associating to each particle of the simulation a Gaussian distribution
with the FWHM equal to the seeing, and in each surface element of the
plane of the sky we compute the surface density of particles $\Sigma$
as the sum of the contributions from the Gaussians and the radial
velocity as the weighted average of the individual velocities. 
In the following, a case corresponding to a particular
projection of a simulation (characterized by wind velocity, wind angle
and time past since the onset of RPS) will be referred to as `case' or
`model', without distinction.

\begin{figure*}
\includegraphics[width=108mm]{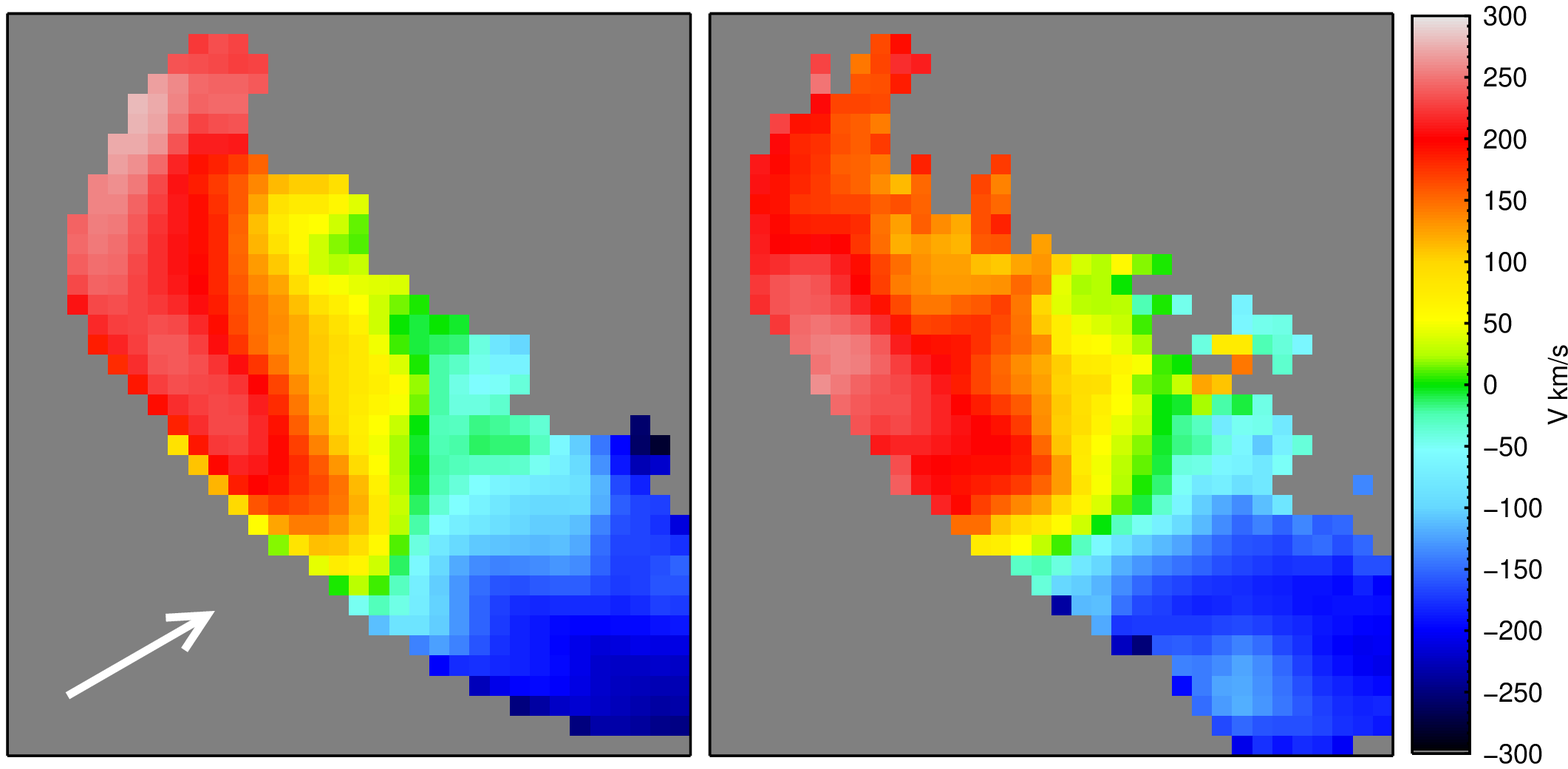} 
\includegraphics[width=52mm]{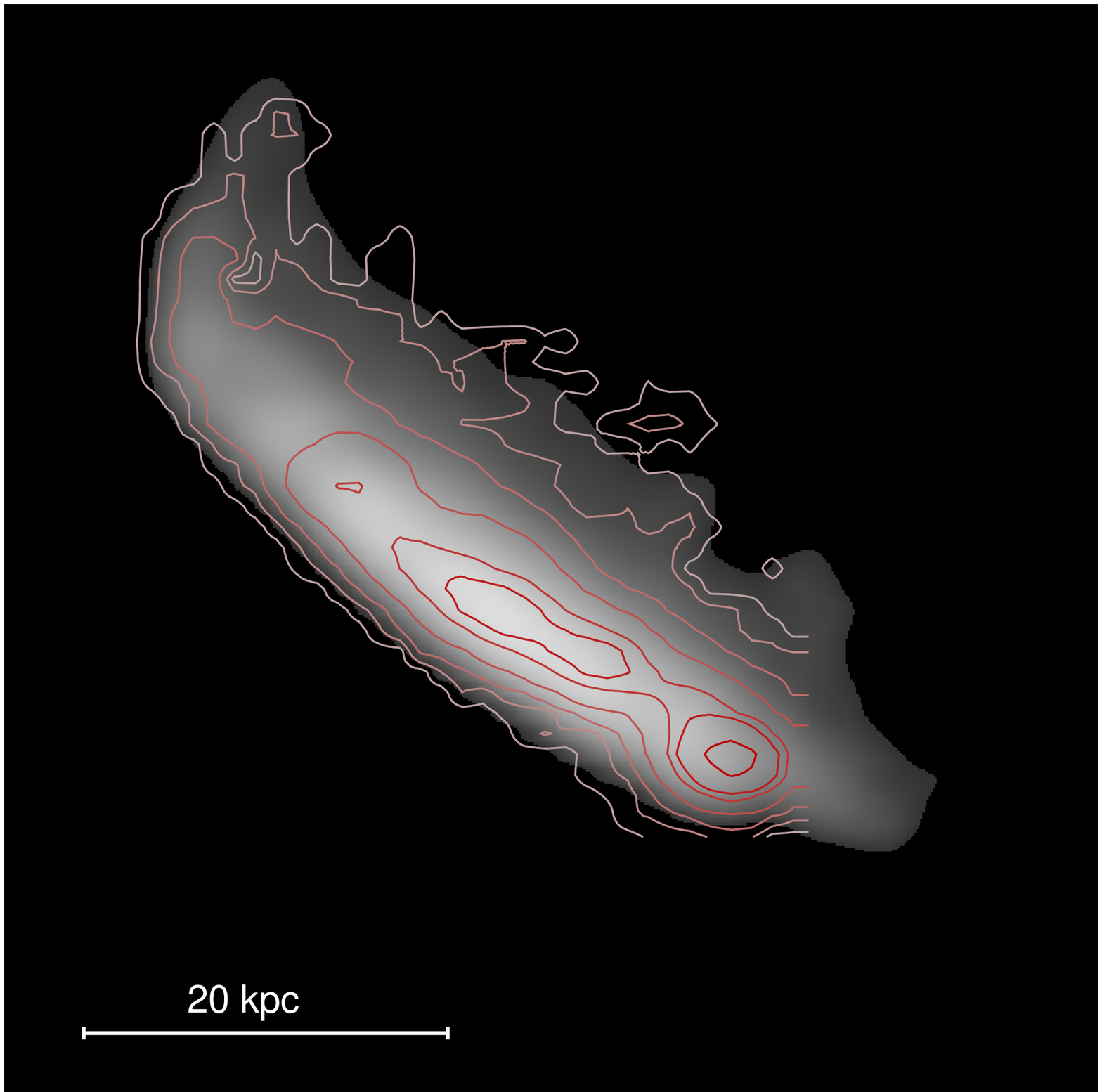}
\caption{{\em Left}: One of the best models ( $\beta=$ 45$^{\circ}$,
$V_{wind}$=1400\,km\,s$^{-1}$, t=60\,Myr; left) is compared to the
observed radial velocity field. The arrow shows the direction of the
ICM wind. The velocity scale is to the right. {\em Right}: The
contours of H$\alpha$ flux measured with WiFeS are superimposed to the
gas density of the same model.}

\label{simul3}
\end{figure*}

For what concerns the spatial distribution of the gas, we should keep in
mind that while the simulations show the velocity field of the whole
gas, our data refer to the ionized gas. However, it is reasonable to
assume that the neutral and ionized gas are mixed. In their study of
the ram-pressure stripped galaxy NGC\,4330, Abramson et
al. (\citeyear{AKC11}) showed that the H$\alpha$ flux is truncated at
the same radius on the SE side as the HI emission which extends
slightly further in the trailing side, and Tonnesen \& Bryan
(\citeyear{TB10}) found that in the tail of the H$\alpha$ emission
follows the neutral gas.

The first selection was made on low-resolution models. To start the
selection, we compared the spatial distributions and the velocity
fields of the models with the observations as explained below.

The H$\alpha$ surface brightness $F_{ij}$ was first normalized to
5$\times$10$^{-18}$erg\,s${-1}$cm$^{-2}$\AA${-1}$arcsec$^{-2}$ so that
at our 5$\sigma$ detection limit $F$=1.  Then, for each case, we
computed the quantity $\chi^2=\sum(F_{ij}-\Sigma_{ij})^2 /
(N_{pix}-1)$, where $\Sigma$ is the particle surface density of the
model and $N_{pix}$ is the number of pixels where $\chi^2$ is
evaluated. The sum was computed, in a first iteration, over the
intersection between the observed area and the area where
$\Sigma>$0. In this computation we masked out the two regions with
very high star-formation rates, i.e. the nuclear region and the region
12 kpc SW from the nucleus. We then ranked the cases in increasing
order of $\chi^2$ and by visual inspection verified that the ranking
in $\chi^2$ reflects faithfully the consistency of the model gas
distribution with the data.

The observed gas is confined in an area limited by our detection limit
of $F$=1, while the $\Sigma$ of the models may assume arbitrarily
small values. To set a lower limit to $\Sigma$ consistent with the
observational limits, we derived, for each case, the mean $a$ of the
ratios $\Sigma_{ij}/F_{ij}$ computed over the flux range $F$=1-10
(i.e. from our flux limit to 10$\times$ higher), and derived a mean
value of $a$ averaging over the first ranked cases. We found
$a$=1.21$\pm$0.23, with the low dispersion reflecting the stability of
$a$ over $\sim$ 100 different models. By definition, $a$ is the
particle surface density corresponding to our lowest detection value,
i.e. to the ``borders'' of our data, and as such it was adopted as the
cutoff value to compare the LR models with the observations. For the
HR simulations, having 10 time higher particle density, we used a
cutoff 10$\times$ higher than this ($a^\prime$=12.1) .

We then run a second iteration of the computation of $\chi^2$, where
we set $\Sigma>a$ to cut the models consistently with the data. We
also considered the size of the intersection between the observed area
and the models, and accepted only the models which cover at least 80\%
of the observed area. In this way we could reject models whose spatial
gas distribution is significantly different from the observed
one. These are the cases when a substantial fraction of the gas is
stripped from the disc, like in advanced phases of RPS (t$>$70 Myr) or
in early phases (t$<$20 Myr), where the gas in the models is still
confined in the disc. Only 38\% cases survived this selection.

A similar ranking was applied to the velocity fields, defining
$\chi^2=\sum(V^m_{ij}-V^o_{ij})^2 / (N_{pix}-1)$, where $V^m_{ij}$ and
$V^o_{ij}$ are the model and observed radial velocity fields
respectively. In this case the ranking according to increasing values
of $\chi^2$ is more prone to `mistakes' than before, and actually the
preferred cases are those at the extremes of time: t$\geq$80 Myr or
t$\leq$20 Myr. The reason is that those models spatially overlap with
the data only in the inner regions, where the velocity field of the
gas is less perturbed. To remove the contamination from these cases,
we had to exclude models with overlap less than 90\%, which left only
5\% of them. Also in the case of velocity fields we verified by visual
inspection that the ranking reflects the consistency between data and
models. The cases that survived this selection were further inspected
visually to look for manifest inconsistencies between models and data.

Figures~\ref{simul1} and \ref{simul2} show some models in two
parameter grids ($\beta$,t and $\beta$,$V_{wind}$ respectively). Some
selected models are marked with a circle on the upper-left corner of
the thumbnails. From the low-resolution simulations, we selected the
models with $\beta$=$45^{\circ}$ and $\beta$=$55^{\circ}$ with
$V_{wind}$=1400\,km\,s$^{-1}$ and 1700\,km\,s$^{-1}$, and disc scale
height 10\% of the scale length, and carried out high-resolution
simulations with these parameters. Lower and higher velocity values at
the same $\beta$ values are not able to produce the observed gas
outflowing or truncate the gas disc at smaller radii than observed,
respectively (see Fig.~\ref{simul2}). At a certain velocity value,
$\beta$ values closer to the edge-on RPS tends to truncate the NE
galaxy side sharply respect to the observations, while $\beta$ values
closer to the face-on RPS presents blueshifted gas in the NW side of
the disc in contrast with what observed (see Figs.~\ref{simul1} and
\ref{fGVF}).

The comparison of the observed velocity field with the high-resolution
simulations leads to a further restriction of the parameter space.
The best final model is shown in Fig.~\ref{simul3} (left panel)
together with the observed velocity field. The projection on the plane
of the sky of the ICM wind is also shown. This model corresponds to
$\beta=45^{\circ}$, $V_{wind}$=1400\,km\,s$^{-1}$ and $t=$ 60\,Myr.
Figure \ref{simul3} (right panel) compares the gas density of the same
model to the observed H$\alpha$ flux distribution. The most external
isophote corresponds to our 5$\sigma$ detection limit ($F=1$) while
the density of the model is consistently cut at $\Sigma=12.1$
particles per arcsec$^2$ (see above). We notice a remarkable agreement
between the model and WiFeS observations. In the southwestern part of
the disc not covered by WiFeS data the model is less constrained by the
H$\alpha$ imaging (cf. Fig.~\ref{halpha} right panel) which is
slightly shallower than the IFS observations.

Our data are therefore consistent with a wind angle $\beta \sim
45^{\circ}$, such that the RPS is taking place at an angle
intermediate between face-on and edge-on. The wind velocity is
1400\,km\,s$^{-1}$ (ram pressure $\sim 9\cdot
10^{12}$dyn\,cm$^{-2}$). The only orientation consistent with the
observations corresponds to the case in which we are observing the
galaxy disc from the side of impact of the ICM wind, in such a way
that the SE side of the disc is the furthest from us. All selected
models correspond to the disc height scale of 0.1$\times r_d$. The
range of the parameters is very narrow. Models with $\beta <
45^{\circ}$ or $\beta > 55^{\circ}$ are clearly less consistent with
the observations, as are the models with $V_{wind} >$
1700\,km\,s$^{-1}$ or $V_{wind} <$ 1400\,km\,s$^{-1}$. For what
concerns the time since the onset of the RPS, the best models are
those at t$\sim $60\,Myr.  We notice that the time scale of the RPS
suggested by the simulations is indicative, but the observational
evidence of the outflowing gas following the rotation of the galaxy
shows that the stripping process has started recently. Considering the
adopted value of the ICM density and the estimated wind velocity we
obtain a ram pressure of 2.5$\cdot 10^{-12}$\,dyn\,cm$^{-2}$.

\section{Summary and Conclusions}
 \label{con}

We have presented IFS observations of the bright spiral galaxy
SOS\,114372, located at 1\,Mpc from the centre of the rich cluster
A\,3558 in the Shapley supercluster core. IFS data demonstrate the
presence of a one-sided ionized gas stream extending 13\,kpc in
projection from the galaxy disc. Complementary narrow-band
H$\alpha$ imaging shows multiple compact knots of H$\alpha$ emission
often connected to faint filamentary strands which extend back to the
galactic disc, while the $K$-band image shows a symmetric
disc. These data, complemented by UV, FIR and deep optical-NIR ($B$,
$R$, $K$) imaging, have allowed us to investigate the gas and stellar
kinematics as well as the properties of the gas and stellar
populations which all together indicate that this galaxy is being
affected by ram-pressure stripping.

The observations have been interpreted by means of emission-line
diagnostics, stellar population synthesis and photoionization
models and N-body/hydrodynamical simulations. The main results of
our analysis are the following.

\begin{description}
\item $i)$ The velocity field of the gas is complex and, although
  still dominated by the potential of the galaxy, is affected by
  significant perturbations even in the inner parts of the disc. The
  velocity dispersion field is characterized by maxima ($\gtrsim$
  100\,km\,s$^{-1}$) localized in the periphery of the disc. The
  velocity dispersion maintains relatively high values ($\gtrsim$
  40\,km\,s$^{-1}$) all over the field. At the same time, the stellar
  velocity field appears to be symmetric with respect to the centre,
  with the typical shape of a barred galaxy.

\item $ii)$ All the gas in the galaxy is contaminated by some fraction of
  shock excitation ($0.05 \lesssim f \lesssim 0.1$) as shown by the
  shock and photoionization models. The galaxy regions associated to
  the outflowing gas and three regions on the opposite side of the
  disc are dominated by shock-excited gas ($0.4 \lesssim f \lesssim
  0.8$), and all but one are strongly attenuated by dust. In these
  regions we also measure the highest gas velocity dispersions.

\item $iii)$ The current star formation rate, measured through the
  H$\alpha$ luminosity, from IFS observations, amounts to
  7.2$\pm$2.2\,M$_{\odot}$yr$^{-1}$ and shows two prominent peaks in
  the centre and at 12 kpc to the SW, which contribute 30\% and 20\%
  to the total SFR, respectively. This picture is fully consistent
  with that measured from the 24$\mu$m flux. The analysis of the
  stellar populations led to the identification a starburst
  characterized by a ${\sim}5{\times}$ increase in the SFR over the
  last $\sim$100\,Myr in the SW disc, a short starburst ($\sim
  300$\,Myr) just quenched in the NE disc and in the galaxy centre a
  predominant population of old ($>$1\,Gyr) stars together with newly
  (${\lesssim}100$\,Myr) formed stars. The intense nuclear star formation
  ($\sim$ 2 M$_\odot$ yr$^{-1}$) is most probably associated with the
  presence of the bar, whose effect is to concentrate the gas towards
  the centre thus triggering star formation \citep[][and references
  therein]{EPP11,OOY12}.

\item$iv)$ Gas flowing out of the discs of galaxies is observed also
in case of tidal interactions or galactic winds. Among the closest
galaxies to \\ SOS\,114372 we do not find any likely companion
responsible for the gas outflow. In fact, SOS\,112392 and SOS\,114493
present regular morphologies with no signs of interaction. The other
candidate for the tidal interaction, the faint galaxy SOS\,115228 with
unknown redshift, shows a starburst which could be related to an
interaction with SOS\,114372, but its very low mass as well as the
relative positions of these two galaxies cannot justify the observed
gas outflow. We also considered the possibility of a galactic wind,
but the morphology of the outflow does not agree with the bipolar
outflows expected in that case. On the other hand, the high velocity
dispersion and the gas moving towards the observer in a triangular
region along the minor axis SE from the nucleus is suggestive of the
semi-cone of a galactic wind produced by the nuclear starburst with
the other semi-cone obscured by dust. In this case the wind would be
also affected by the RPS and dispersed on a short time scale.

\end{description}

The RPS scenario for SOS\,114372 is also supported by {\it ad hoc}
{\it N}-body/hydrodynamical simulations. They allowed us to constrain
the angle between the rotation axis and the ICM wind,
$\beta\sim45^{\circ}$, the wind velocity, $V_{wind} \sim
1400$\,km\,s$^{-1}$, and the ``age'' of the gas stripping,
t$\sim$60\,Myr. The fact that the ram-pressure stripping started
60\,Myr ago is consistent with the burst of star formation in the SW
disc starting less than 100\,Myr ago according to the stellar
population modeling. We are thus led to ascribe this starburst to the
compression of the ISM caused by the ram pressure.

In this work we have analyzed the case of an L* galaxy affected by
ram-pressure stripping. It is the most luminous (M$_B\sim$-21.5) and
distant ($z_{A3558}$=0.048) galaxy showing {\it ongoing} RPS studied
in detail (c.f. galaxies in Virgo, Coma and A\,1367 clusters;
e.g. Chung et al. \citeyear{CvG09} and reference therein; Abramson et
al. \citeyear{AKC11}; Vollmer et al. \citeyear{VSC09}; Yagi et
al. \citeyear{YYK10}). SOS\,114372 is located far away from the
cluster centre, in an environment characterised by a relatively low
density with respect to the cluster's core. This adds a piece of
evidence to the fact that RPS is acting more efficiently than
previously foreseen and also outside of the cluster cores as also
observed in the Virgo cluster by \citet{CGK07}. This new understanding
of the RPS candidates supports the view that this mechanism is the
principal transformation process to turn spirals into S0s \citep{KB12}
although helped by other processes affecting the structure of the
galaxies.

We plan to analyse the supercluster sample observed with WiFeS in
forthcoming papers providing other examples of the influence of the
environment on galaxy evolution.

\section*{Acknowledgments}
This work was carried out in the framework of the collaboration of the
FP7-PEOPLE-IRSES-2008 project ACCESS. CPH acknowledges financial
support from STFC. RJS was supported under STFC rolling grand
PP/C501568/1 ``Extragalactic Astronomy and Cosmology at Durham
2008-2013''. Dopita acknowledges the support from the Australian
Department of Science and Education (DEST) Systemic Infrastructure
Initiative grant and from an Australian Research Council (ARC) Large
Equipment Infrastructure Fund (LIEF) grant LE0775546 which together
made possible the construction of the WiFeS instrument. Dopita would
also like to thank the Australian Research Council (ARC) for support
under Discovery project DP0984657. DS and SS acknowledge the
UniInfrastrukturprogramm des BMWF Forschungsprojekt Konsortium
Hochleistungsrechnen and the Austrian Science Foundation FWF through
grant W1227 (FWF DK-CIM) and thanks Volker Springel for providing the
simulation code GADGET-2 and A. B\"ohm for profitable discussions. The
authors wish to thank the staff at the Australian National University
2.3m telescope for their assistance for WiFeS observations. PM, GB and
CPH thank the ANU-RSAA for hospitality. PM and GB thank C.~J. Farage
for helpful discussions about the use of WiFes pipeline. CPH thanks
Sylvain Veilleux for his assistance in the acquisition and calibration
of the Magellan/MMTF H$\alpha$ imaging and the technical staff at the
Las Campanas Observatory for their support during the observations. A
special thanks to A.~J.~R. Sanderson for providing the X-ray profile
of A\,3558 and to K.A. Pimblett and A.~W. Graham for their comments
which helped to improve the manuscript. Figures 1 and 5 were prepared
with the IDL code provided by Lupton et al. (2004) and available at
http://cosmo.nyu.edu/hogg/visualization/rgb/. The authors thank the
referee B. Vollmer for his constructive comments and suggestions.

\appendix
\section[]{Exploiting the simulations}
\label{App}

To compare the results of simulations with the data, the simulated
data must be first correctly oriented with respect to the observer and
then processed to mimic the characteristics of the observing
instrument. These two steps are described in the next subsections.

\subsection{Observing the simulations}

The output of the hydrodynamical simulations are tables with one row
for each particle containing the three components of position and
velocity. In our case the models are oriented with the spatial
velocity of the galaxy $\overrightarrow{V}_{gal}$ along the (positive)
Z-axis and the Y-axis on the galaxy disc. We first apply a rotation
around the Y-axis to move the galaxy disc in the (X,Y) plane, so that
$\overrightarrow{V}_{gal}$ lies in the positive XZ semi-plane.
Fig.~\ref{appenfig1} represents a coordinate system in which the disc
of the galaxy lies in the (X,Y) plane. The direction to the observer
is marked by the dashed lines ending with a small circle. Two viewing
situations, differing by a rotation of 180$^{\circ}$ around the
X-axis, are in principle possible. In one case (case {\it A}) the wind
points towards Z$>$0 (red arrows in the figure), while in the other
(case {\it B}) the wind points towards Z$<$0 (cyan arrow).

To project the model into the plane of the sky, we require that the
spatial velocity of the galaxy projects into the observed
line-of-sight velocity. The angle $\phi$ between
$\overrightarrow{V}_{gal}$ and the line of sight is given by $cos\
\phi = (V_{los}/V_{gal})$. This is equivalent to require that
$\overrightarrow{V}_{wind}$ makes an angle $\phi$ with the line of
sight in the direction of the observer.

The top panels of Fig.~\ref{appenfig1} show the case where
$\beta=$25$^{\circ}$ and V$_{wind}$=2400\,km\,s$^{-1}$, or
$\phi\sim$70$^{\circ}$. The cone with $\beta$=const. (hereafter `$\beta$-cone';
 blue) intersects the cone $\phi$=const. (hereafter `$\phi$-cone'; 
black/gray) in two points in the Z$>$0 semi-sphere. The two
black lines are the projection of the semi-planes passing through the
Z-axis and the intersection points of the cones. They are at angles
$\theta_1$ and $\theta_2$ ($-32^{\circ}$, $212^{\circ}$ in this
example) with the X-axis. Rotating the galaxy by one of these two
angles, moves the wind direction at the right angle with respect to
the line of sight in which it projects into $V_{los}$. In this case
there are therefore two possible directions from which we can
``observe'' our model. In both views, the wind is directed as the red
arrow in the upper-right panel, which corresponds to case {\it
A}. Since the wind cone does not intersect the $\phi$-cone in the
Z$<$0 hemisphere, no other viewing angles are possible.

\begin{figure}
\includegraphics[width=84mm]{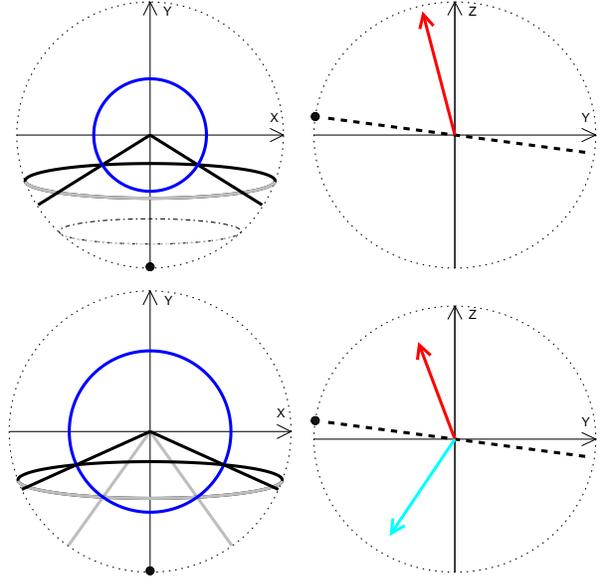}
\caption{Unit sphere (outline dotted), centred on the galaxy centre,
with the X and Y axes on the galactic plane, and the Z-axis along the
symmetry axis of the disc. The direction to the observer is marked by
the small black filled circle in the semi-plane Y$<$0, Z$>$0, and is
inclined by i=82$^{\circ}$ with respect to the Z-axis. In this way,
the X-axis corresponds to the apparent major axis of the galaxy. The
upper panels represent the case $\beta$=25$^{\circ}$ and
V$_{wind}$=2400\,km\,s$^{-1}$, while the lower panels represent the case
$\beta$=45$^{\circ}$ with the same V$_{wind}$. This wind speed
corresponds to $\phi$=69.64$^{\circ}$. The left panels offer a view
of the unit sphere from the Z-axis. The blue circles are the
intersection with the sphere of the cones $\beta$=const. The black and
gray ellipses are the intersection with the unit sphere of the semi-cone
$\phi$=const. directed towards the observer. The gray section is the
half of the ellipse laying at Z$<$0, i.e. at the opposite side of the
galaxy disc in this view. The right panels show the same
configurations projected on the YZ plane. The dashed line is the line
of sight. The dot-dashed ellipse in the upper-left panel is the
intersection of the unit sphere with the cone corresponding to
V$_{wind}$=1100\,km\,s$^{-1}$, an example of a forbidden configuration.
The arrows represent projections of the ICM wind (see text).}
\label{appenfig1}
\end{figure}

In the bottom panels of the same figure (corresponding to the case
$\beta=$45$^{\circ}$ and $\phi\sim$70$^{\circ}$) the wind cone
intersects the $\phi$-cone in both hemispheres. There are
intersections in the Z$>$0 hemisphere at $\theta_1$,$\theta_2$
(=-24.3$^{\circ}$, 204.3$^{\circ}$; black lines), and for Z$<$0 the
intersections take place at $\theta_3$,$\theta_4$ (=-54.4$^{\circ}$,
234.4$^{\circ}$ gray lines). Rotating the model by $\theta_1$ or
$\theta_2$, allows us, as before, to see the model from the correct point
of view. But in this case the model can also be rotated around the
Z-axis by $\theta_3$ or $\theta_4$.

The second case corresponds to the cyan wind vector in the
bottom-right panel and represents an example of case {\it B}. 
To reach this configuration, the model should be first rotated by
180$^{\circ}$ around the X-axis.

The coordinate transformations needed in the above process are the
following. Let $\beta$ be the angle of the wind with respect to the
symmetry axis of the galaxy and $i$ the inclination angle of the
galaxy (plane of the sky to galaxy disc). First we move the symmetry
axis of the galaxy into the Z-axis with a clock-wise rotation of
$\beta$ around the Y-axis. In this way, the wind lies on the (X,Z)
plane. The second step is to move the wind
vector in a direction in which the wind speed projects into
$\overrightarrow{V}_{gal}$. It can be shown that his is achieved by a
rotation around the Z-axis by an angle $\theta$ given by

$$cos\  \theta = - \frac{cos\  \phi + k\  cos\  \beta\  cos\ i}{sin\
  \beta\  sin\  i}
\ \ \ ,$$

\noindent
where $k=-1$ for case {\it A} and $k=1$ for case {\it B}. For each
$k$ this equation has two solutions: $\theta_1 = arcos (cos\ \theta)$
and $\theta_2 = \pi - \theta_1$. As a result, when only case {\it A}
is allowed, there are two possible viewing angles (or `projections'),
while when both cases {\it A} and {\it B} are allowed, the possible
projections are four.

Other transformations are possible that do not change the physics of
the models, but given that we fixed the position of the line of sight
in the YZ plane, the only remaining meaningful transformation is a
reflection with respect to that plane. This corresponds to an
inversion of the X-axis and to the change of sign of the corresponding
velocity component of the model particles. This transformation makes
the projection of the models on the sky to reflect with respect to the
apparent major axis, and may become useful to match the models to the
observed orientation of the galaxy, when this is not achievable with a
simple rotation. In the present case this transformation was not
needed.

Cases {\it A} and {\it B} represent two quite different observing
conditions. In case {\it A}, the wind acts on the disc side opposite
to the observer, in such a way that the outflowing gas overlays the
disc. In case {\it B} the opposite is true, and part of the
outflowing gas is hidden by the disc.

Finally, we notice that not all combinations of parameters lead to an
observable configuration. An example is illustrated by the dot-dashed
ellipse in the top-left panel of Fig.~\ref{appenfig1}, which is the
intersection with the unit sphere of a $\phi$-cone corresponding to
V$_{wind}$=1100\,km\,s$^{-1}$. This cone has no intersection with the
$\beta$-cone corresponding to $\beta$=25$^{\circ}$. This means that
there is no possible orientation such that this configuration would
lead to the observed line-of-sight velocity of our galaxy.

\subsection{Simulating the observations}

The end product of the previous process is a matrix
(x$_i$,y$_i$,V$_i$; i=1,...,N$_{particles}$), where (x,y) are
coordinates in the plane of the sky (x is along the apparent major
axis), y is normal to x, V is the radial velocity and N$_{particles}$
is the number of particles in the simulation.

To reproduce the effect of the seeing we substituted each particle by
a Gaussian distribution with the FWHM equal to the seeing. In each
surface element of the plane of the sky (see below), we computed the
surface density of particles $\Sigma$ as the sum of the contributions
from the Gaussians and the radial velocity as the weighted average of
the individual velocities.

To speed-up the computation, we truncated the contribution of each particle
at 10$\times \sigma_{Gauss}$. This computation
was performed on a grid of points spaced by 0.1${\prime\prime}$ on
the sky, which we verified to offer a reasonable compromise between
accuracy and computation speed. The points in the plane of the sky
were eventually sampled in $1{\prime\prime}\times 1{\prime\prime}$
bins, matching the spatial binning of the data, and rotated to the
position angle of the major axis of our galaxy.

\end{twocolumn}
\label{lastpage}
\end{document}